\documentclass[preprint,aps,amsmath,nofootinbib,tightenlines]{revtex4}
\usepackage{graphicx}
\usepackage{bm}

\newcommand{\nn}{\nonumber}

\newcommand{\bqa}{\begin{eqnarray}}
\newcommand{\eqa}{\end{eqnarray}}

\usepackage{ifpdf}
\ifpdf \DeclareGraphicsExtensions{.pdf, .jpg}

\else
\DeclareGraphicsExtensions{.eps, .jpg,.ps}

\fi

\begin{document}

\title{Exclusive Double Charmonium Production from $\Upsilon$ Decay}

\author{Yu Jia\footnote{Electronic address: jiay@ihep.ac.cn}}
\affiliation{Institute of High Energy Physics, Chinese Academy of
Sciences, Beijing 100049, China\vspace{0.2cm}}

\date{\today\\ \vspace{1cm} }



\begin{abstract}

The exclusive decay of $\Upsilon$ to a vector plus pseudoscalar
charmonium is studied in perturbative QCD.  The corresponding
branching ratios are predicted to be of order $10^{-6}$ for first
three $\Upsilon$ resonances, and we expect these decay modes
should be discovered in the prospective high-luminosity $e^+e^-$
facilities such as super $B$ experiment.  As a manifestation of
the short-distance loop contribution,  the relative phases among
strong, electromagnetic and radiative decay amplitudes can be
deduced.  It is particularly interesting to find that the relative
phase between strong and electromagnetic amplitudes is nearly
orthogonal.  The resonance-continuum interference effect for
double charmonium production near various $\Upsilon$ resonances in
$e^+e^-$ annihilation is addressed.

\end{abstract}

\maketitle

\newpage

\section{Introduction}

Very rich $J/\psi$ decay phenomena have historically served as an
invaluable laboratory to enrich our understanding toward the
interplay between perturbative and nonperturbative
QCD~\cite{Kopke:1988cs,Brambilla:2004wf}. By contrast,  much fewer
decay channels of $\Upsilon$ are known to date.  It would be
definitely desirable if more knowledge about bottomonium decay can
be gleaned.

The typical branching fraction for a given hadronic decay mode of
$\Upsilon$ is in general much smaller than that of $J/\psi$. It is
partly due to the smaller QCD coupling  at the $b$ mass scale than
at the $c$ scale, and more importantly, it is because the
branching ratio gets diluted by a scaling factor of $(m_c/m_b)^n$
when descending from charmonium to bottomonium (here $n$ is some
number no less than 4).  These might intuitively explain why very
few exclusive decay modes of bottomonia have been seen so far.

Due to rather large $b$ mass, $\Upsilon$ not only can
dematerialize into light hadrons,  it also can decay to charmful
final states. In this work,  we plan to study the exclusive decay
of $\Upsilon$ into double charmonium,  or more specifically,
$J/\psi$($\psi^\prime$) plus $\eta_c$($\eta_c^\prime$). The hard
scales set by $b$ and $c$ masses in this type of processes justify
the use of perturbative QCD (pQCD). Since the involved mesons are
all heavy quarkonium,  it is natural to employ NRQCD factorization
approach~\cite{Bodwin:1994jh}.  This work constitutes a
continuation of previous studies on bottomonium decay to double
charmonium, namely, $\chi_b,\,\eta_b\to
J/\psi\,J/\psi$~\cite{Braguta:2005gw,Jia:2006rx}.  Although these
decay modes have not yet been seen,  some experimental information
have already been available for inclusive $J/\psi$ ($\psi^\prime$)
production rate from $\Upsilon$
decay~\cite{Maschmann:1989ai,Abe:2001za,Briere:2004ug}:
\bqa
 {\cal B}[\Upsilon(1S) \to J/\psi+X]&=& (6.5\pm 0.7)\times
10^{-4}\,, \nn \\
 {\cal B}[\Upsilon(1S) \to \psi^\prime+X]&=& (2.7\pm 0.9)\times
10^{-4}\,, \nn \\
{\cal B}[\Upsilon(2S) \to J/\psi+X]&<& 6\times 10^{-3}\,, \nn \\
{\cal B}[\Upsilon(4S) \to J/\psi+X]&<& 1.9\times 10^{-4}\,.
\label{inclusive:Psi:production:ups} \eqa
These inclusive decay ratios set upper bounds for our exclusive
processes. It is worth noting that $\Upsilon\to J/\psi\eta_c$
violates the hadron helicity
conservation~\cite{Brodsky:1981kj,Chernyak:1983ej}. It is thus
natural to expect that the corresponding branching fractions are
very suppressed.

One important impetus of this work is from the double charmonium
production at $\Upsilon(4S)$ resonance measured by Belle in
2002~\cite{Abe:2002rb}.   The observed cross section is usually
entirely  ascribable to the continuum contribution because of
rather broad $\Upsilon(4S)$ width.  Nevertheless for a full
understanding,  it is worth knowing precisely the impact of the
resonant decay on the measured double charmonium cross section.
Furthermore, stimulated by Belle's discovery, a natural question
may arise--what is the discovery potential for double charmonium
production in $e^+e^-$ experiments operated at lower $\Upsilon$
peaks?  Since the first three $\Upsilon$ resonances are much
narrower than $\Upsilon(4S)$,  the resonant decay contribution
should dominate over the continuum one. Our study is motivated to
answer this question.

One interesting problem in exclusive decays of a vector quarkonium
is to know the relative phase between strong and electromagnetic
amplitudes.  For example, the corresponding relative phase in
$J/\psi\to PV$ ($P$, $V$ stand for light $0^{-+}$ and $1^{--}$
mesons) has been extensively studied and found to be nearly
orthogonal~\cite{Baltrusaitis:1984rz,Coffman:1988ve,Jousset:1988ni,
LopezCastro:1994xw,Suzuki:1998ea,Achasov:2001wy}. In our case, the
relative phase naturally emerges as a short-distance effect and
thus is perturbatively calculable.  Curiously, it is also found to
be approximately orthogonal.

The rest of the paper is organized as follows. In
Section~\ref{CSM:Calculation}, we present the lowest-order NRQCD
calculation for the decay process $\Upsilon\to J/\psi+\eta_c$,
including strong, electromagnetic and radiative decay channels. In
Section~\ref{phenomenology},  we present the predictions to the
branching fractions for various $\Upsilon$ decays to double
charmonium,  and conclude that the discovery potential of these
decay modes is promising in the prospective Super $B$ experiment.
We also discuss the relative phases among three amplitudes,
putting particular emphasis  on the nearly orthogonal relative
phase between strong and electromagnetic amplitudes. The
connection between our results and the previous discussions on the
nearly $90^\circ$ relative phase in $J/\psi$ decays is remarked.
In addition,  we also study the impact of the resonance-continuum
interference on $J/\psi+\eta_c$ production cross sections  at
various $\Upsilon$ resonances in $e^+e^-$ experiments.  We
summarize and give a brief outlook in Section~\ref{summary}. In
the Appendixes, we illustrate how to analytically derive some loop
integrals that appear in Section~\ref{CSM:Calculation}.

\section{Color-singlet Model Calculation}
\label{CSM:Calculation}

The process $\Upsilon\to J/\psi+\eta_c$ can proceed via three
stages: the $b\bar{b}$ pair first annihilates into three gluons,
or two gluons plus a photon, or a single photon; in the second
step, these highly virtual gluons/photon then convert into two
$c\bar{c}$ pairs, which finally materialize into two fast-moving
$S$-wave charmonium states.  Due to the heavy charm and even much
heavier bottom, both the annihilation of $b\bar{b}$ and creations
of $c\bar{c}$ pairs take place in rather short distances,  it is
thereby appropriate to utilize pQCD to study this hard exclusive
process.

This process is somewhat similar to the widely studied $J/\psi\to
PV$ decay, but bears the virtue that applicability of pQCD should
be more reliable. It is commonly believed that some
nonperturbative mechanisms should play a dominate role in many
charmonium exclusive decay processes,  where the credence of  pQCD
seems rather questionable. This  consensus  is exemplified  by the
notorious $\rho\pi$ puzzle~\cite{Brambilla:2004wf,Mo:2006cy}.

While it is customary to use the light-cone approach to deal with
hard exclusive processes involving light mesons (for a recent
attempt to study $J/\psi\to \rho\pi$ from this perspective, see
Ref.~\cite{Li:2007pb}),  it is for our purpose most proper to
employ an approach embodying the non-relativistic nature of
quarkonium.  {\it NRQCD factorization} is a widely accepted
effective-field-theory framework to describe the quarkonium
inclusive production and decay processes,  which incorporates
systematically the small velocity expansion~\cite{Bodwin:1994jh}.
Although a rigorous formulation for exclusive quarkonium decay has
not yet been fully achieved within this scheme,  one may still be
well motivated to work with models akin to the NRQCD ansatz.

The {\it color-singlet model} can be viewed as a truncated version
of NRQCD approach, in which one still assumes a factorization
formula, {\it i.e.}, the decay rate can be separated into the
perturbatively calculable part and  universal nonperturbative
factors, however only with the contribution from the color-single
channel retained. We do not know how to include the possible
color-octet contributions in a clear-cut way,  but it is plausible
to assume their effects are unimportant for reactions involving
only $S$-wave quarkonium as in our case.  Notice that NRQCD and
color-singlet model are often referring to the same tool in
literature, so we will also use them interchangeably.

Let $Q$, $P$ and $\tilde{P}$ signify the momenta of $\Upsilon$,
$J/\psi$ and $\eta_c$, respectively.  In color-singlet model
calculation, one starts with the parton process
$b(p_b)\,\bar{b}(p_{\bar b}) \to c(p_c) \,\bar{c}(p_{\bar{c}}) +
c(\tilde{p}_c) \,\bar{c}(\tilde{p}_{\bar{c}})$, then projects
this matrix element onto the corresponding color-singlet quarkonium
Fock states.
This work is intended only for the zeroth order in
relativistic expansion,  hence we can neglect the
relative momenta inside each quarkonium, {\it i.e.}, set
$p_b=p_{\overline b}=Q/2$, $p_c=p_{\overline c}=P/2$, and
$\tilde{p}_c=\tilde{p}_{\overline c}=\tilde{P}/2$.  For the $ b
\bar{b}$ pair to be in  a spin-triplet and color-singlet state,
one simply replaces the product of the Dirac and color spinors
for $b$ and $\overline{b}$ by
the projection operator
\bqa
u(p_b)\,\overline{v}(p_{\bar b})& \longrightarrow& {1\over 2
\sqrt{2}} \,(\not\! Q+ 2 m_b)\,\not\! \varepsilon_{\Upsilon}
\times \left( {1\over \sqrt{m_b}} \psi_{\Upsilon}(0)\right)
\otimes {{\bf 1}_c\over \sqrt{N_c}}\,.
\label{Upsilon:projector} \eqa
For the outgoing $J/\psi$ and $\eta_c$, one makes the following
replacements:
\bqa
v(p_{\bar c})\,\overline{u}(p_c)& \longrightarrow& {1\over 2
\sqrt{2}} \not\! \varepsilon^*_{J/\psi}\,(\not\! P+2 m_c)\, \times
\left( {1\over \sqrt{m_c}} \psi_{J/\psi}(0)\right) \otimes
{{\bf 1}_c\over \sqrt{N_c}}\,, \label{JPsi:projector}
\\
v(\tilde{p}_{\bar c})\,\overline{u}(\tilde{p}_c)& \longrightarrow&
{1\over 2 \sqrt{2}} \, i\gamma_5\,(\not\! \tilde{P}+2 m_c)\, \times
\left( {1\over \sqrt{m_c}} \psi_{\eta_c}(0)\right) \otimes {{\bf
1}_c\over \sqrt{N_c}}\,. \label{Etac:projector} \eqa
Here $\varepsilon^{\mu}_{\Upsilon}$ and
$\varepsilon^{\mu}_{J/\psi}$ are polarization vectors for
$\Upsilon$ and $J/\psi$.  $N_c=3$, and ${\bf 1}_c$ stands for the
unit color matrix. The nonperturbative factors
$\psi_{\Upsilon}(0)$,  $\psi_{J/\psi}(0)$ and $\psi_{\eta_c}(0)$
are Schr\"{o}dinger wave functions at the origin
for $\Upsilon$, $J/\psi$ and $\eta_c$,
which can be inferred either from phenomenological potential models
or extracted from experiments. By writing (\ref{Upsilon:projector}),
(\ref{JPsi:projector}) and (\ref{Etac:projector}) the
way as they are, it is understood that $M_{\Upsilon}=2m_b$ and
$M_{J/\psi}\approx M_{\eta_c} =2m_c$ have been assumed.

Before moving into the concrete calculation, we recall first that
since both strong and electromagnetic interactions conserve
parity,  the decay amplitude is then constrained to have the
following  Lorentz structure:
\bqa
{\cal M} &=& {\cal A} \:\epsilon_{\mu \nu \alpha \beta}\,
\varepsilon^\mu_{\Upsilon}\,
\varepsilon^{*\nu}_{J/\psi}\,Q^\alpha\,P^\beta\,.
\label{Lorentz:tensor:structure}
\eqa
Apparently, $J/\psi$ must be transversely polarized in $\Upsilon$
rest frame.  All the dynamics is encoded in the coefficient ${\cal
A}$, which we call {\it reduced} amplitude.  Our task in the
remaining section then is to dig out its explicit form.

\subsection{Three-gluon Amplitude}

\begin{figure}[tb]
\begin{center}
\includegraphics[scale=0.7]{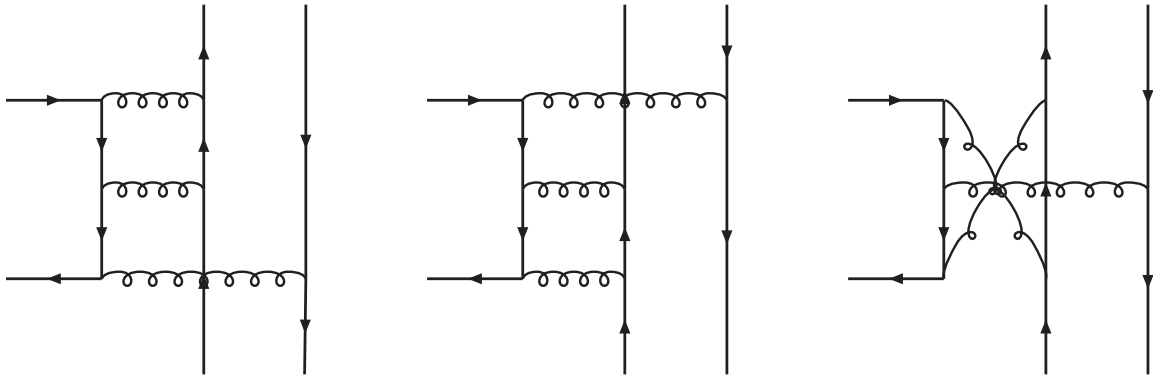}
\caption{Some representative lowest-order diagrams that contribute
to $\Upsilon \to 3g\to J/\psi+\eta_c$. \label{feynman:diag:3g}}
\end{center}
\end{figure}

We begin with the strong decay amplitude.  Some typical
lowest-order diagrams are shown in Fig.~\ref{feynman:diag:3g},
which starts already at one loop order.  Using the projection
operators in (\ref{Upsilon:projector}), (\ref{JPsi:projector}),
and (\ref{Etac:projector}), we can write down the corresponding
amplitude:
\bqa {\cal M}_{3g}&=&  2\,N_c^{-3/2}\,{\rm tr}(T^a T^b T^c)\,{\rm
tr}(T^a \{T^b,T^c\} )\, g_s^6\,
{\psi_{\Upsilon}(0)\,\psi_{J/\psi}(0)\,\psi_{\eta_c}(0)\over 16
\sqrt{2}\, m_b^{7/2} m_c} \int\!\! {d^4k_1 \over
(2\pi)^4}\,{1\over k_1^2}\, {1\over k_2^2}
\nn \\
& \times & \left\{ {{\rm tr}[(\not\! Q+ 2m_b) \not\!
\varepsilon_{\Upsilon} \gamma^\rho\gamma^\nu (\not \! k_2 +m_b)
\gamma^\mu] \over k_2^2-m_b^2 } \right. + {{\rm tr}[(\not\! Q+
2m_b) \not\! \varepsilon_{\Upsilon} \gamma^\nu (-\not \! k_1 +m_b)
\gamma^\mu\gamma^\rho] \over k_1^2-m_b^2 }
\nn\\
&- & \left. { m_b\,{\rm tr} [\not\! Q+ 2m_b) \not\!
\varepsilon_{\Upsilon} \gamma^\mu (-\not \! k_2+m_b)\gamma^\rho
(\not\! k_1+m_b)\gamma^\nu  ] \over (k_1^2-m_b^2)\,(k_2^2-m_b^2)}
\right\}
\nn \\
& \times &  { {\rm tr}[\not\! \varepsilon^*_{J/\psi} (\not\!
P+2m_c) \gamma_\mu (\not\!p_c - \not\!k_1 + m_c) \gamma_\nu
\gamma_5(\not\!\tilde{P}+ 2m_c)\gamma_\rho] \over
(p_c-k_1)^2-m_c^2 } \,, \label{QCD:ampl:Feyn:Rule} \eqa
where  two internal gluons carry momenta $k_1$ and $k_2$,
respectively, which are subject to the constraint $k_1+k_2={Q\over
2}$. Some elaboration is in order.  Because $\Upsilon$ has charge
conjugation quantum number $-1$,  three intermediate gluons must
arrange to the color-singlet state
$d^{abc}|a\rangle|b\rangle|c\rangle$ [$d^{abc}$($f^{abc}$)
represents the totally (anti)symmetric structure constants of
$SU(N_c)$ group]. This restriction removes all the possible ${\cal
O}(g_s^6)$ diagrams involving 3-gluon vertex. As a result, we only
need retain those Abelian diagrams in which each of three gluons
is connected between the $b$ and $c$ quark lines in both ends.
There are totally twelve such diagrams, but it turns out that for
each of diagrams, there is another one generating exactly
identical amplitude,  which explains the prefator 2 in the right
hand side of (\ref{QCD:ampl:Feyn:Rule}). Among the six diagrams
needed to be considered,   one can further divide them into two
groups: one carries a color factor   $\propto {\rm tr}(T^a T^b
T^c){\rm tr}(T^a T^c T^b)$, whereas the other carries that
$\propto {\rm tr}(T^a T^b T^c){\rm tr}(T^a T^b T^c)$. These two
groups yield identical reduced amplitudes  except this difference.
Thus we only need consider three diagrams with distinct
topologies,  as depicted in Fig.~\ref{feynman:diag:3g}, and
incorporate the following color factor:
\bqa
{\rm tr}(T^a T^b T^c)\,{\rm tr}(T^a \{T^b,T^c\} )&=& {1\over 8}\,
d_{abc} d^{abc}  = {(N_c^2-1)(N_c^2-4) \over 8 N_c} \,, \eqa
which reassures us that only those intermediate gluons with
overall $C=-1$ can contribute to this process.

Straightforward power counting reveals that the loop integrals in
(\ref{QCD:ampl:Feyn:Rule}) are  simultaneously  ultraviolet and
infrared finite.  In absence of the need for regularization, we
have directly put the spacetime dimension to four.

After completing the Dirac trace in (\ref{QCD:ampl:Feyn:Rule}), we
end up with terms in which the Levi-Civita tensor is entangled
with the loop momentum variable.  Since  all these terms will
finally conspire to arrive at the desired Lorentz structure as
dictated in (\ref{Lorentz:tensor:structure}),  we may exploit this
knowledge to get rid of the antisymmetric tensor prior to
performing the loop integral~\cite{Guberina:1980xb}. First we may
identify the partial amplitude $M_{\mu\nu}$ through ${\cal M}
=M_{\mu\nu}\,\varepsilon^\mu_\Upsilon\,
\varepsilon^{*\nu}_{J/\psi}$. Equation
(\ref{Lorentz:tensor:structure}) then demands
\bqa
M_{\mu\nu} &=& {\cal A} \:\epsilon_{\mu \nu \alpha \beta}\,
Q^\alpha\,P^\beta\,,
\label{red:amp:mu:nu}
\eqa
Contracting both sides of (\ref{red:amp:mu:nu}) with
$\epsilon^{\mu \nu \rho\sigma} Q_\rho P_\sigma$, one can extract
the reduced amplitude using
\bqa {\cal A}  &= & {1\over 2M_\Upsilon^2 |{\bf P}|^2}\,
\epsilon^{\mu \nu \rho\sigma}  M_{\mu\nu} Q_\rho P_\sigma\,,
\label{red:am:shortcut} \eqa
where $|{\bf P}|=[(Q\cdot P)^2-Q^2 P^2]^{1/2}/M_\Upsilon$ is the
modulus of the momentum of $J/\psi$ ($\eta_c$) in the $\Upsilon$
rest frame.

After this manipulation is done,  we end in a concise expression
\bqa
{\cal A}_{3g} &=& {2\sqrt{2}\, (N_c^2-1)\,(N_c^2-4)\over
N_c^{5/2}}{\pi\, \alpha_s^3 \over \,m_b^{7/2} |{\bf P}|^2} \,
\psi_{\Upsilon}(0) \psi_{J/\psi}(0)
\psi_{\eta_c}(0)\,f\left({m_c^2\over m_b^2}\right) \,,
 \label{red:am:3g}
\eqa
where $f= f_1+f_2+f_3$, and
\bqa f_1 &=&  \int\! {d^4 k_1\over i\pi^2}\, {
(m_b^2-4m_c^2)(k_2^2-m_b^2) + k_1\cdot(3 Q-P)\,k_1\cdot
P-(1+m_c^2/m^2_b) (k_1\cdot Q)^2 \over k_1^2\,k_2^2\,
(k_2^2-m_b^2)\,(k_1^2 - k_1\cdot P)} \,,\label{3g:f1}
\\
f_2 &=&  \int\! {d^4 k_1 \over i\pi^2}\, {(m_b^2-4m_c^2)
(k_1^2-m_b^2) + k_2\cdot P \,k_2\cdot \tilde{P} - (m_c^2/m_b^2)
(k_2\cdot Q)^2 \over k_1^2\,(k_1^2-m_b^2)\,k_2^2\,(k_1^2 -
k_1\cdot P)} \,, \label{3g:f2}
\\
f_3 &=& m_b^2 \int\! {d^4 k_1 \over i\pi^2} \,{k_1\cdot(Q-2P)
(k_1^2 - k_1\cdot\tilde{P})- 2(m_b^2-4m_c^2)k_1\cdot k_2 \over
k_1^2\,(k_1^2-m_b^2)\,k_2^2\,(k_2^2-m_b^2)\, (k_1^2 - k_1\cdot
P)}\,. \label{3g:f3} \eqa
Since $f_i$ is dimensionless, it can depend upon $m_b$ and $m_c$
only through their dimensionless ratio $m_c^2/ m_b^2$. These loop
integrals can be worked out analytically, and the results are
\bqa {\rm Re}\, f(\xi) &=& 3-{2 \,\pi\over \sqrt{3}}
+4(1-2\xi)\left\{{1\over 1-\beta}\ln\left[{1+\beta\over 2}\right]+
{1\over 1+\beta}\ln\left[{1-\beta\over 2}\right]\right\}
\nn \\
&-& 2(1+2\xi) \left\{ {1\over (1-\beta)^2}\ln\left[{1+\beta\over
2}\right]+ {1\over (1+\beta)^2}\ln\left[{1-\beta\over 2}\right]
+{1\over 4 \xi}\right\}
\nn \\
&-&  {1-2\xi\over \beta} \left\{2 \,{\rm tanh}^{-1}\beta\,\ln \xi
+ 2\, {\rm Li}_2\left[{1-\beta\over 2}\right] -2\, {\rm
Li}_2\left[{1+\beta\over 2}\right] + {\rm Li}_2\left[{\beta-1\over
\beta+1}\right]\right.
\nn \\
&-& \left. {\rm Li}_2\left[ {\beta+1\over \beta-1}\right] \right\}
+{4 \xi\over \beta} \left\{ {2 \pi \over 3}{\rm
tan}^{-1}[\sqrt{3}\beta] + 2\,{\rm tanh}^{-1}\beta\, \ln [1-3 \xi]
 \right.
\nn \\
&+& {\rm Li}_2\left[{2\,\beta\over 1+\beta}\right] -{\rm
Li}_2\left[{2\,\beta\over \beta-1}\right] + {\rm Li}_2\left[{\beta
(\beta+1)\over \beta-1}\right] - {\rm
Li}_2\left[{\beta(1-\beta)\over 1+\beta}\right]
\nn \\
&+&{\rm Li}_2\left[{\beta(1+\beta)\over 2 (1-3 \xi)}\right] -{\rm
Li}_2\left[{\beta(\beta-1)\over 2 (1-3 \xi)}\right] + {\rm
Li}_2\left[-{\beta(1-\beta)^2\over 4 (1-3 \xi)}\right]
 -{\rm
Li}_2\left[{\beta(1+\beta)^2\over 4 (1-3 \xi)}\right]
\nn \\
&+& \left. 2\,{\rm Re}\left\{ {\rm Li}_2\left[-{(1+
i\,\sqrt{3})\,\beta \over 1-i\,\sqrt{3}\,\beta } \right]-{\rm
Li}_2\left[{(1+ i\,\sqrt{3})\,\beta \over 1+i\,\sqrt{3}\,\beta }
\right]\right\} \right\}\,,
\label{re:f:tot}\\
 {\rm Im}\, f(\xi) &=& \pi \left\{ 1-{2\,(1-2 \xi) \,{\rm tanh}^{-1}\beta
\over \beta}\right\}\,,
\label{im:f:tot}
 \eqa
where ${\rm Li}_2$ is the dilogarithm function, and
$\beta=\sqrt{1-4 \xi}$.  We will illustrate in
Appendix~\ref{f:analytic} how to obtain this result. The emergence
of imaginative part of $f$ characterizes the contribution from two
on-shell internal gluons.  The shapes of the real and imaginary
parts of $f$ are displayed in Fig.~\ref{plot:f:g}.

It is instructive to know the asymptotic behavior of $f$ in the
$\xi\to 0$ limit. This can be readily read out from
(\ref{re:f:tot}) and (\ref{im:f:tot}),
\bqa {\rm Re}f(\xi)&=& {1\over 2}\ln^2 \xi +{3\over 2}\ln \xi
+1+{\pi^2\over 6} - {2\,\pi\over \sqrt{3}}+ {\cal O}(\xi\ln\xi)
\,,\label{Re:f:asymptotic}
\\
{\rm Im}f(\xi)& = &  \pi\,(\ln \xi+1)+ {\cal O}(\xi\ln\xi)
\,.\label{Im:f:asymptotic} \eqa
Note both the real and imaginary parts blow up logarithmically in
the limit $\xi \to 0 $, as can be clearly visualized in
Fig.~\ref{plot:f:g}. These (quadratically) logarithmical
divergences in the $m_c\to 0$ limit are obviously of infrared
origin. Nevertheless, this does not pose any practical problem,
since a nonrelativistic description for a zero-mass bound state,
as well as the resulting predictions, should not be trusted
anyway.  It is interesting to note that, provided that $\xi$ is
not overly small, say, $\xi>10^{-4}$,  then $-{\rm Im}\,f$ is
always bigger than $|{\rm Re}\, f|$,  or more precisely phrased,
$-{3\pi\over 4}<{\rm arg}\,f<-{\pi\over 4}$.

\subsection{Two-gluon-one-photon Amplitude}

\begin{figure}[tb]
\begin{center}
\includegraphics[scale=0.72]{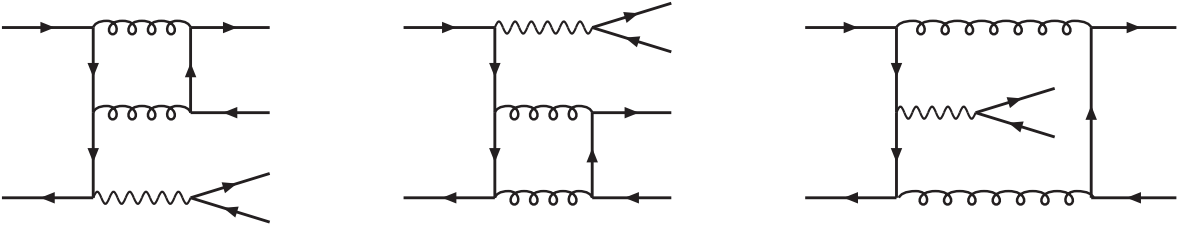}
\caption{Three representative lowest-order diagrams that
contribute to $\Upsilon \to gg \gamma \to J/\psi+\eta_c$, where
the $J/\psi$ comes from the photon fragmentation.
\label{feynman:diag:gggamma}}
\end{center}
\end{figure}

We next turn to the contribution through the radiative decay
channel. $C$-parity conservation demands that one end of photon
line must be attached to the $b$ quark. Those diagrams obtained
from replacing one gluon by one photon in
Fig.~\ref{feynman:diag:3g} do contribute, however their magnitudes
are much less important than the diagrams shown in
Fig.~\ref{feynman:diag:gggamma}, which essentially proceed as
$\Upsilon\to gg(\to \eta_c)+\gamma(\to J/\psi)$.   This is because
in the latter case, the $J/\psi$ is created via the photon
fragmentation,  which thereby receives a $m_b^2/m_c^2$ enhancement
relative to the former.  We will only consider the latter case, in
which the lowest-order contribution also starts at one loop. Using
the projection operators in (\ref{Upsilon:projector}),
(\ref{JPsi:projector}), and (\ref{Etac:projector}), it is
straightforward to write down the corresponding amplitude:
\bqa {\cal M}_{gg\gamma}&=&  2\,N_c^{-1/2}\,{\rm tr}(T^a
T^b)\,{\rm tr}(T^a T^b)\,e_b e_c e^2 g_s^4\,{\psi_{\Upsilon}(0)\,
\psi_{J/\psi}(0)\,\psi_{\eta_c}(0)\over 8\sqrt{2}\,m_b^{1/2}
m_c^2} \int\!\! {d^4k_1 \over (2\pi)^4}\,{1\over k_1^2}\, {1\over
k_2^2}
\nn \\
 & \times & \left\{ {{\rm tr}[(\not\!
Q+ 2m_b) \not\! \varepsilon_{\Upsilon} \not\!
\varepsilon^*_{J/\psi} (\not \! P- \not\! p_b +m_b) \gamma^\nu
(\not \! p_b- \not\! k_1 +m_b)\gamma^\mu] \over ((P-p_b)^2-m_b^2)
((p_b-k_1)^2-m_b^2) } \right.
\nn \\
&+& {{\rm tr}[(\not\! Q+ 2m_b) \not\! \varepsilon_{\Upsilon}
\gamma^\nu (\not \! k_2- \not\! p_b +m_b) \gamma^\mu (\not \! p_b-
\not\! P +m_b) \not\! \varepsilon^*_{J/\psi} ] \over
((P-p_b)^2-m_b^2) ((p_b-k_2)^2-m_b^2) }
\nn\\
&+& \left. {{\rm tr}[(\not\! Q+ 2m_b) \not\!
\varepsilon_{\Upsilon}  \gamma^\nu  (\not \! k_2- \not\! p_b +m_b)
\not\! \varepsilon^*_{J/\psi} (\not \! p_b- \not\! k_1 +m_b)
\gamma^\mu] \over ((p_b-k_1)^2-m_b^2) ((p_b-k_2)^2-m_b^2) }
\right\}
\nn \\
& \times &  { {\rm tr}[\gamma_5(\not\!\tilde{P}+ 2m_c) \gamma_\mu
(\not\!\tilde{p}_c - \not\!k_1 + m_c) \gamma_\nu ] \over
(\tilde{p}_c-k_1)^2-m_c^2 } \,, \label{QCD:amp2:Feyn:Rule} \eqa
where the momenta carried by two internal gluons are labelled by
$k_1$, $k_2$, which satisfy $k_1+k_2=\tilde{P}$. The factor 2 in
the right side of (\ref{QCD:amp2:Feyn:Rule}) takes into account
the identical contributions from other three crossed diagrams.

\begin{figure}[tb]
\begin{center}
\includegraphics[scale=0.6]{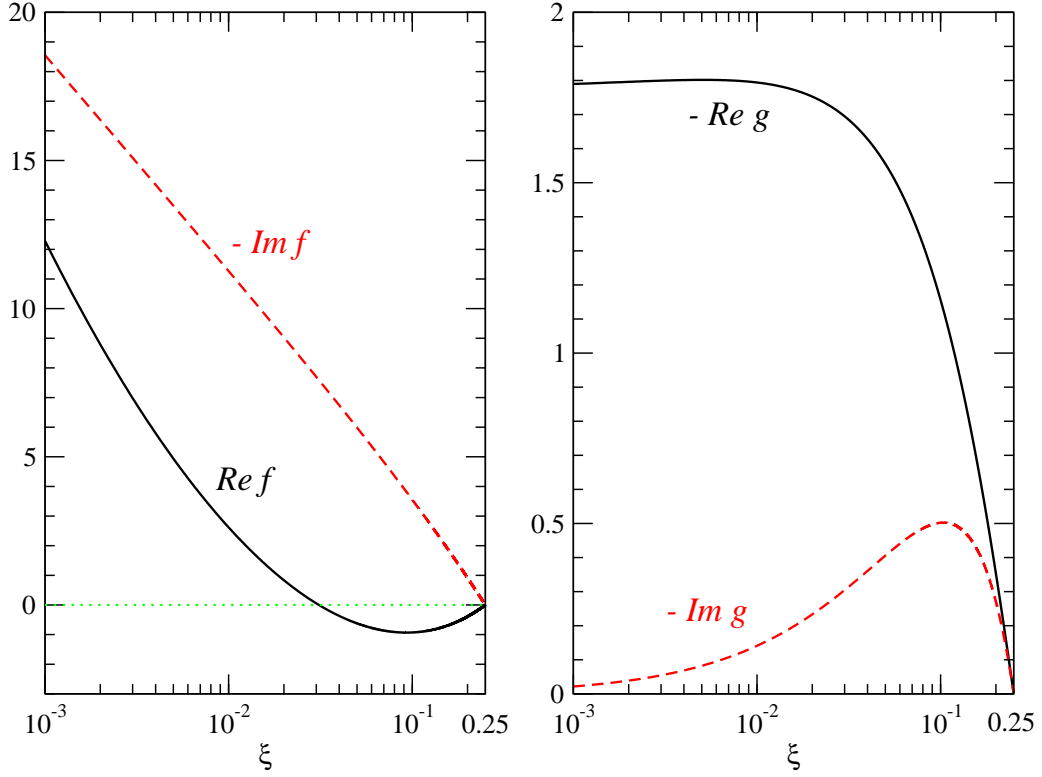}
\caption{Real and imaginary parts of $f(\xi)$ and $g(\xi)$.
\label{plot:f:g}}
\end{center}
\end{figure}

Following the same shortcut adopted in the $3g$ channel, we can
derive the desired reduced amplitude with recourse to
Eq.~(\ref{red:am:shortcut}),
\bqa {\cal A}_{gg\gamma} &=& {4 \sqrt{2}\,(N_c^2-1) \over
N_c^{1/2}} {e_b e_c \pi \alpha \alpha_s^2 \,m_b^{1/2} \over
(m_b^2- 2 m_c^2)  |{\bf P}|^2\,
m_c^2}\,\psi_{\Upsilon}(0)\psi_{J/\psi}(0)\psi_{\eta_c}(0)\,
g\left({m_c^2\over m_b^2}\right) \,, \label{red:am:2g:gamma}
 \eqa
where the dimensionless function $g$ is defined by
\bqa g\left({m_c^2\over m_b^2}\right) &=& \int\! {d^4 k_1\over
i\pi^2}\, { (2 m_c^2/m_b^2\,Q\cdot r - P\cdot r )\,\tilde{P}\cdot
r + 2 (m_b^2-4m_c^2)\, r^2  \over k_1^2 k_2^2 (k_1^2-k_1\cdot Q)
(k_2^2 - k_2\cdot Q)} \,. \label{gggamma:def:g} \eqa
For convenience,  we have introduced a new internal momentum
variable $r$, which is defined through $k_1=\tilde{P}/2+r$ and
$k_2=\tilde{P}/2-r$.  Note that the integrand is symmetric under
$r\to -r$,  reflecting the symmetry $k_1\leftrightarrow k_2$.  A
gratifying fact is that the charm propagator has now been
cancelled in the denominator.  We dedicate
Appendix~\ref{g:analytic} to a detailed derivation of this loop
integral.  Like its counterpart $f$ in the three-gluon channel,
the function $g$ is both ultraviolet and infrared finite.  Its
analytic expression reads
\bqa {\rm Re}\, g(\xi)&=& (1-2 \xi)\ln[2-4 \xi]+ 4\sqrt{\xi(1-\xi)}\,
\tan^{-1}\sqrt{\xi \over 1-\xi}
\nn \\
&-&   \xi \beta \left\{ 4\,{\rm tanh}^{-1}\beta\,\ln [2 \xi] +2\,{\rm
Li}_2[-\beta] -2\,{\rm Li}_2[\beta]+{\rm Li}_2\left[{\beta-1\over
\beta+1}\right] \right.
\nn \\
&-& \left. {\rm Li}_2\left[{\beta+1\over \beta-1}\right]+ {\rm
Li}_2\left[{2\,\beta\over (1+\beta)^2}\right] - {\rm
Li}_2\left[-{2\,\beta\over (1-\beta)^2}\right] \right\}\nn \\
 &-& {(1-2 \xi)^2\over \beta} \left\{ {\rm Li}_2[\beta]- {\rm
Li}_2[-\beta]+2\,{\rm Re} \left\{ {\rm
Li}_2\left[{(1+\beta)^2+4i\sqrt{\xi(1-\xi)}\over 4(1-2 \xi)}\right]
\right.\right. \nn \\
&-& {\rm Li}_2\left[{(1-\beta)^2+4i\sqrt{\xi(1-\xi)}\over
4(1-2 \xi)}\right]+{\rm
Li}_2\left[-{\beta(1-\beta)^2+4i\beta\sqrt{\xi(1-\xi)}\over
4(1-2 \xi)}\right]  \nn \\
&-& \left.\left. {\rm
Li}_2\left[{\beta(1+\beta)^2+4i\beta\sqrt{\xi(1-\xi)}\over
4(1-2\xi)}\right] \right\}\right\}\,, \label{re:g:tot}
\\
{\rm Im} \,g(\xi)&=& -2 \pi \,\xi \beta \,{\rm tanh}^{-1}\beta \,.
\label{im:g:tot} \eqa

The shapes of real and imaginary parts of $g$ are displayed in
Fig.~\ref{plot:f:g}. Note that $-{\rm Re}\,g$ is always bigger
than $-{\rm Im}\, g$ for any $\xi$,  or put in another way,
$-\pi<{\rm arg}\,g<- {3\pi\over 4}$.  Apparently,  the imaginary
part of $g$ vanishes as $\xi\to 0$, whereas the real part of $g$
approaches the following asymptotic value:
\bqa {\rm Re}\, g(\xi)&=&  -{\pi^2\over 4}+ \ln 2 + {\cal
O}(\xi\ln\xi)\,. \eqa
In contrast to $f$,  both of the real and imaginary parts of $g$
admit a finite value in the $\xi\to 0$ limit.  Not surprisingly,
the asymptotic behavior of this function is quite similar to the
analogous one in the $\Upsilon\to \eta_c\gamma$
process~\cite{Guberina:1980xb}.

\subsection{Single-photon Amplitude}

\begin{figure}[tb]
\begin{center}
\includegraphics[scale=0.6]{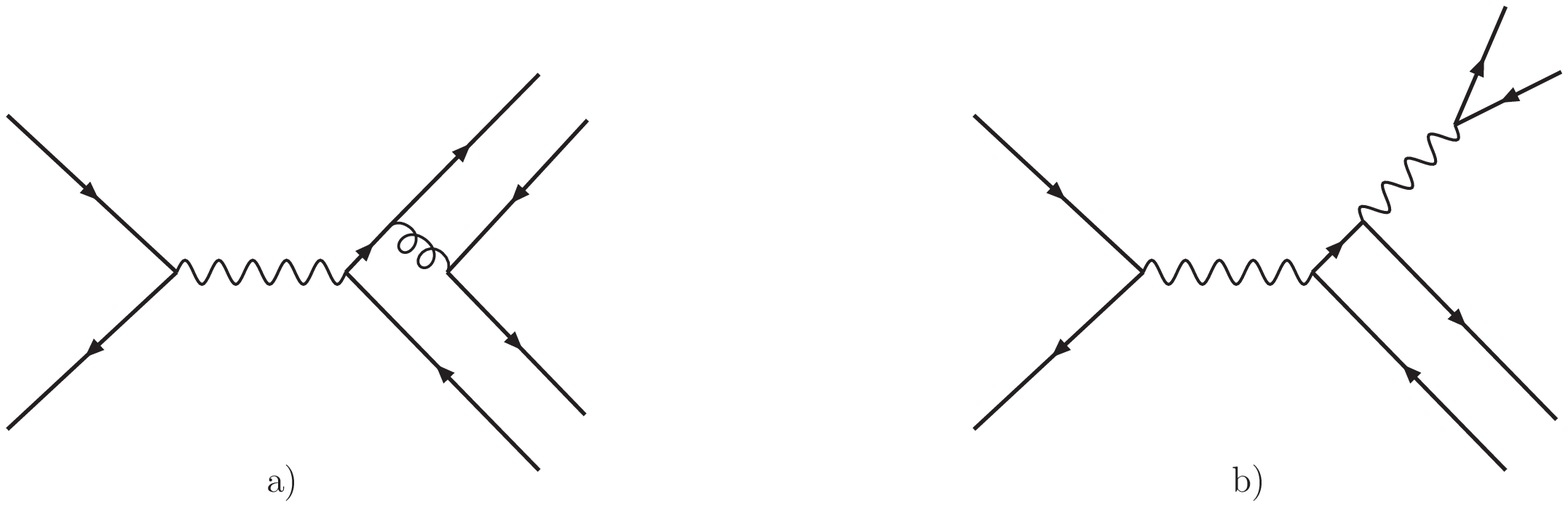}
\caption{Two representative lowest-order diagrams that contribute
to $\Upsilon \to \gamma^* \to J/\psi+\eta_c$.  There are totally
four diagrams in class a) and two in class b).
\label{feynman:diag:gamma}}
\end{center}
\end{figure}

Let us now consider the electromagnetic contribution via the
annihilation of $b\bar{b}$ into a single photon,  with some
typical diagrams shown in Fig.~\ref{feynman:diag:gamma}.  This
process is closely related to the contiuum $J/\psi+\eta_c$
production in $e^+e^-$ annihilation,  which has recently aroused
much attention since the measurements were first released  by
Belle collaboration~\cite{Abe:2002rb}. Rather unexpectedly, it is
shortly found that the leading-order NRQCD prediction to the
production cross section falls short of the data by about one
order of magnitude~\cite{Braaten:2002fi,Liu:2002wq}, which
subsequently triggered intensive theoretical efforts to resolve
this alarming
discrepancy~\cite{Bodwin:2002fk,Bodwin:2002kk,Hagiwara:2003cw,Ma:2004qf,
Bondar:2004sv,Zhang:2005ch,Bodwin:2006dm,He:2007te}.

In the Born order,  one can directly import the time-like
electromagnetic form factor of $S$-wave charmonium first deduced
in Ref.~\cite{Braaten:2002fi} to here, and the corresponding
lowest-order one-photon amplitude reads
\bqa
{\cal A}_{\gamma}
 &=&
 -{16\sqrt{2} (N_c^2-1)\over {N_c}^{1/2}} {\pi^2 e_b e_c \alpha\,\alpha_s
 \over m_b^{11/2}} \psi_{\Upsilon}(0) \psi_{J/\psi}(0) \psi_{\eta_c}(0)
 \left(1+{N_c^2\over 2(N_c^2-1)}\,
 {e_c^2 \alpha \,m_b^2\over \alpha_s \,m_c^2}\right),
 \label{red:am:gamma*}
\eqa
where the second term in the parenthesis represents the pure QED
contribution in which $J/\psi$ arises from photon fragmentation,
as is represented by Fig.~\ref{feynman:diag:gamma}b).

Recent calculations indicate that the $J/\psi+\eta_c$
electromagnetic form factor is subject to large perturbative and
relativistic corrections at $B$ factory
energy~\cite{Zhang:2005ch,He:2007te}. It seems that the disturbing
discrepancy between $B$ factories measurements and NRQCD
predictions have been largely resolved once these large
corrections are taken into account.   Motivated by this, from now
on we will replace the entities in the parenthesis in
(\ref{red:am:gamma*}) by a positive constant $K$ ($>1$),  which
presumedly encompasses all the radiative and relativistic
corrections.

\subsection{Decay Width and Asymptotic Scaling Behavior}

It is now the time to lump  three different contributions
together.  Plugging (\ref{red:am:3g}), (\ref{red:am:2g:gamma}),
and (\ref{red:am:gamma*}) into the formula
\bqa \Gamma[\Upsilon \to J/\psi+\eta_c ] &=& { |{\bf P}|^3 \over
12\,\pi}\,|{\cal A}_{\gamma}+{\cal A}_{3g}+{\cal
A}_{gg\gamma}|^2\,, \eqa
we then obtain the desired decay partial width.  Note the cubic
power of momentum reflects that $J/\psi$ and $\eta_c$ are in
relative $P$ wave.  This formula has already taken into account
the spin average of  $\Upsilon$ and the polarization sum over
$J/\psi$. The result is
\bqa \Gamma[\Upsilon\to J/\psi+\eta_c] &=& \Gamma[\Upsilon\to e^+
e^-] \nn \\
&\times&  {2^{20} \,\pi^2 e_c^2 \,\alpha_s^2\, |{\bf P}|^3\over
9\, M^9_\Upsilon} \,\psi^2_{J/\psi}(0) \,\psi^2_{\eta_c}(0)
|a_\gamma+ a_{3g}+ a_{gg\gamma}|^2, \label{Partial:Width} \eqa
where $a_\gamma = K$,
\bqa a_{3g}&=& -{5 \,\alpha_s^2 \over 72 \pi\, e_b e_c \,\alpha}\,
{m_b^2\over |{\bf P}|^2 } \,f\left({m_c^2\over m_b^2}\right)\,,
\label{a3g:analytic}
\\
a_{gg\gamma} &=& - {\alpha_s \over 4 \pi}\,{m_b^6 \over
(m_b^2-2m_c^2) |{\bf P}|^2\,m_c^2}\, g\left({m_c^2\over
m_b^2}\right) \,, \label{agggamma:analytic} \eqa
and $\Gamma[\Upsilon\to e^+e^-]= 16\pi e_b^2\alpha^2
\psi^2_\Upsilon(0)/M_\Upsilon^2$  is the electronic width of
$\Upsilon$.

It is instructive to deduce the asymptotic behaviors of these
three different contributions. Because we are more concerned about
the power-law scaling,  we  will take $f, g\sim {\cal O}(1)$ for
simplicity since they vary with quark masses logarithmically at
most.   Assuming $\psi_{J/\psi}(0)\sim \psi_{\eta_c}(0)\sim (m_c
v_c)^{3/2}$ ($v_c$ is the typical relative velocity between $c$
and $\bar{c}$),  from (\ref{Partial:Width}) we find
\bqa {\Gamma[\Upsilon\to\gamma^*\to J/\psi+\eta_c] \over
\Gamma[\Upsilon\to e^+ e^-]} & \sim &  \alpha_s^2\, {m_c^6\over
m_b^6}\, v_c^6\,, \label{scaling:bv:gamma}
\\
{\Gamma[\Upsilon\to 3g\to  J/\psi+\eta_c] \over \Gamma[\Upsilon\to
e^+ e^-]} &\sim&  {\alpha_s^6\over \alpha^2} \, {m_c^6\over
m_b^6}\, v_c^6\,,\label{scaling:bv:ggg}
\\
{\Gamma[\Upsilon\to gg\gamma \to J/\psi+\eta_c] \over
\Gamma[\Upsilon\to e^+ e^-]} &\sim&  \alpha_s^4 \,{m_c^2\over
m_b^2} \, v_c^6\,. \label{scaling:bv:gggamma} \eqa
First interesting  observation is that both
(\ref{scaling:bv:gamma}) and (\ref{scaling:bv:ggg}) exhibit the
$1/m_b^6$ scaling behavior.  This is as expected from the
celebrated helicity selection rule in perturabative QCD, which is
applicable for both single-photon and three-gluon
processes~\cite{Brodsky:1981kj}.  The reason is as follows. The
final-state $J/\psi$ must be transversely polarized, in line with
the parity and Lorentz invariance,  the hadron helicity
conservation $\lambda_{J/\psi}+\lambda_{\eta_c}=0$ is violated by
one unit, hence the ratio is suppressed by an extra $1/m_b^2$
relative to the leading-twist $1/m_b^4$ scaling. In contrast, the
corresponding ratio in $gg\gamma$ channel,
(\ref{scaling:bv:gggamma}), though suppressed by coupling
constants with respect to other two subprocesses, nevertheless
enjoys a much milder ($\sim 1/m_b^2$) kinematical suppression,
because the $J/\psi$ directly comes from the photon fragmentation.
Simple power counting implies that these three different
contributions have comparable strengths for the physical masses of
$b$ and $c$.

Another noteworthy fact is that, there are relative phases among
three amplitudes,  which are encoded in the $f$ and $g$ functions.
Since all these phases originate from the loop integrals,  we may
regard them of short-distance origin.

\section{Phenomenology}

\label{phenomenology}

\subsection{Determination of $K$ from $B$ factories measurement}

\begin{table}[tb]
\caption{Experimental inputs for $\Upsilon(nS)$ and $S$-wave
charmonia (taken from Ref.~\cite{Yao:2006px}).  The last column
lists the wave functions at the origin for various $S$-wave
charmonium states,  retrieved from the measured electric width
through (\ref{jpsi:ee:width:nlo}) by assuming $m_c=1.5$ GeV and
$\alpha_s(2m_c)=0.26$.} \label{exper:input:bbbar:ccbar}
\begin{center}
\begin{tabular}{ccc ccc ccc ccc ccc ccc}
\hline  & $H$  && & Mass  (GeV) & && $\Gamma_{\rm tot}$ (keV) & &&
$\Gamma_{e^+e^-}$ (keV) & && $\psi_H(0)$ (${\rm GeV}^{3/2}$) &
 \\ \hline
& $\Upsilon(1S)$    &&  & $9.460$ & &&
 $54.02\pm1.25$ & &&$1.340\pm 0.018 $  & && -- &
 \\ \hline & $\Upsilon(2S)$    &&  & 10.023& && $31.98\pm
2.63$ & &&
$0.612\pm 0.011 $& &&  -- &  \\
\hline & $\Upsilon(3S)$    &&  & 10.355& && $20.32\pm 1.85$ & &&
$0.443\pm 0.008 $ & && -- &
\\ \hline & $\Upsilon(4S)$    &&  & $10.579$& && $ 20500\pm 2500$ & && $0.272\pm 0.029$
& && -- &
\\ \hline\hline
& $J/\psi$    &&  & $3.097$ & &&
 -- & &&$5.55\pm 0.14$  & && 0.263 &  \\
  \hline
  & $\eta_c$    &&  & $2.980$& &&
  -- & && --  & && 0.263 & \\ \hline
 & $\psi^\prime$ && & $3.686$& && -- & && $2.48\pm0.06$& && 0.176 &
  \\ \hline
  & $\eta_c^\prime$ && & $3.638$& &&
 -- & && --  & && 0.176 &\\ \hline
\end{tabular}
\end{center}
\end{table}

First we want to determine the value of $K$ in
(\ref{Partial:Width}), which characterizes the magnitude of
higher-order corrections to the single-photon amplitude.  For the
sake of simplicity, we will assume the $K$ factors are equal in
our case and in $J/\psi+\eta_c$ production through $e^+e^-$
annihilation to a virtual photon. Of course, this is just an
approximation, because the virtual gluon line connecting $b$ quark
and final-state $c$ quark, as well as the relativistic correction
in $\Upsilon$, which will emerge in our process accounting for the
radiative and relative corrections, are absent in the double
charmonium production in continuum. We will assume these
additional corrections are insignificant.

First we recall the continuum double charmonium  cross section in
the lowest order in $\alpha_s$ and
$v_c^2$~\cite{Braaten:2002fi,Liu:2002wq}:
\bqa \sigma_{\rm cont}[e^+e^- \to J/\psi+\eta_c] &=&
\sigma_{\mu^+\mu^-} \,{2^{20}\pi^2\,e_c^2\alpha_s^2 \over 9 }
{|{\bf P}|^3 \over s^{9/2}} \, \psi^2_{J/\psi}(0)
\psi^2_{\eta_c}(0) \,, \label{double:charmonim:conti:tree} \eqa
where $\sigma_{\mu^+\mu^-} = {4\pi\alpha^2\over 3s}$.  For
simplicity, the pure QED contribution where $J/\psi$ is produced
via photon fragmentation (the analogous diagram to
Fig.~\ref{feynman:diag:gamma}b) has been neglected.

In this work, we extract the wave functions at the origin for
vector charmonium states from their measured electric widths. We
will use the formula incorporating the first order perturbative
correction
\bqa \Gamma[J/\psi\to e^+e^-]&=& {4\pi e_c^2\alpha^2 \over m_c^2}
\psi^2_{J/\psi}(0)\left(1-{8\,\alpha_s(2m_c)\over
3\pi}\right)^2\,. \label{jpsi:ee:width:nlo} \eqa
Heavy quark spin symmetry is then invoked to infer the wave
functions at origin for the corresponding $^1S_0$ charmonium
states.  All the involved charmonium wave functions at origin are
tabulated in Table~\ref{exper:input:bbbar:ccbar}.

If we choose $m_c=1.5$ GeV,  $\alpha_s=0.22$,  we then obtain from
(\ref{double:charmonim:conti:tree}) the tree level continuum
$J/\psi+\eta_c$ cross section at $\sqrt{s}=10.58$ GeV  to be
$4.74$ fb.  This theoretical prediction can be contrasted with the
most recent $B$ factories
measurements~\cite{Abe:2004ww,Aubert:2005tj}:
 \bqa \sigma[e^+e^-\to J/\psi+\eta_c]\times {\cal
B}^{\eta_c}_{> 2} &=& 25.6\pm 2.8 ({\rm stat})\pm 3.4 ({\rm
syst})\;{\rm fb}\,,\hspace{1.2 cm}{\rm Belle}
\nn \\
 \sigma[e^+e^-\to J/\psi+\eta_c]\times {\cal
B}^{\eta_c}_{>2} &=& 17.6\pm 2.8({\rm stat})^{+1.5}_{-2.1}({\rm
syst})\; {\rm fb}\,, \hspace{1.2 cm}B\!\!A\!B\!A\!R
\label{Babar:doub:charmon:meas}
\eqa
where ${\cal B}^{\eta_c}_{> 2}$ represents the branching ratio of
$\eta_c$ decay to more than 2 charged tracks, hence should be less
than 1.   With large uncertainties, both measurements seem to be
marginally consistent with each other.

If we assume the measured  $\sigma_{\rm cont}[e^+e^-\to
J/\psi+\eta_c]$ to be $23$ fb,   and expect that large radiative
and relativistic corrections to
(\ref{double:charmonim:conti:tree}) can bring the leading-order
NRQCD prediction to this value,  we then require
$K=\sqrt{23/4.74}\approx 2.2$. This $K$ factor is roughly
consistent with what is obtained through actual higher-order NRQCD
calculations~\cite{Zhang:2005ch,He:2007te}.  Although we extract
this constant through the $\Upsilon(4S)\to J/\psi+\eta_c$ process,
we will assume it is universal in all other double charmonium
decay channels of $\Upsilon(nS)$.

\subsection{Exclusive decay of $\Upsilon(nS)$ to double $S$-wave charmonium}
\label{excl:dec:ups:2swave:charmonium}

To date, $\Upsilon$ exclusive decays to double charmonium have not
yet been experimentally established.  To make concrete predictions
from (\ref{Partial:Width}),  we need specify the values of all the
input parameters.  We fix $m_c$ to be $1.5$ GeV,  but take $m_b$
as a variable-- for each $\Upsilon(nS)$ decay process, we
approximate it as  half of $M_{\Upsilon(nS)}$. The magnitude of
$|\bf P|$ is determined by physical kinematics. We assume $K=2.2$
for all decay channels, and take the values of the wave functions
at the origin for various charmonium from
Table~\ref{exper:input:bbbar:ccbar}.  As for the coupling
constants, we take $\alpha=1/137$, and $\alpha_s(m_b)=0.22$.  The
uncertainties of our predictions are estimated by sliding the
renormalization scale from $2m_b$ to $m_b/2$ (corresponding to
varying $\alpha_s$ from 0.18 to 0.26). It should be cautioned that
the ambiguity of the inputted  $b$ mass, especially for higher
$\Upsilon$ excitations,  can bring even more severe uncertainty
due to the higher powers of $m_b$ appearing in
(\ref{Partial:Width}).

\begin{table}[tb]
\caption{Predicted  partial widths and branching ratios for
various decay channels of $\Upsilon(nS)$ to vector plus
pseudoscalar charmonium.}
\label{width:B:Upsilon:double:charmonium} {\small
\begin{center}
\begin{tabular}{rrrcccccc|rrrcccccc}
\hline  & Decay channels  && & $\Gamma$ (eV) & && $\cal B$ & &&
Decay channels & && $\Gamma$ (eV)& && $\cal B$&
\\
\hline & $\Upsilon(1S)\to J/\psi+\eta_c$    &&  &
$0.208^{+0.302}_{-0.126}$ & &&
 $3.9^{+5.6}_{-2.3}\times 10^{-6}$ & &&$\Upsilon(2S)\to J/\psi+\eta_c$  & && $0.082^{+0.119}_{-0.050}$ & &&
$2.6^{+3.7}_{-1.6}\times 10^{-6}$ & \\ \hline & $\Upsilon(1S)\to
J/\psi+\eta_c^\prime$    && & $0.109^{+0.185}_{-0.074}$ & &&
$2.0^{+3.4}_{-1.4}\times 10^{-6}$  & &&
$\Upsilon(2S)\to J/\psi+\eta_c^\prime$& &&  $0.042^{+0.067}_{-0.027}$ & && $1.3^{+2.1}_{-0.9}\times 10^{-6}$ & \\
\hline & $\Upsilon(1S)\to \psi^\prime+\eta_c$    &&  &
$0.093^{+0.127}_{-0.054}$& && $1.7^{+2.4}_{-1.0}\times 10^{-6}$ &
&&$\Upsilon(2S)\to \psi^\prime+\eta_c$ & &&
$0.037^{+0.051}_{-0.022}$ & && $1.1^{+1.6}_{-0.7}\times 10^{-6}$ &
\\ \hline & $\Upsilon(1S)\to \psi^\prime+\eta_c^\prime$    &&  & $0.045^{+0.073}_{-0.030}$&
&& $0.8^{+1.4}_{-0.6}\times 10^{-6}$ & &&$\Upsilon(2S)\to
\psi^\prime+\eta_c^\prime$ & && $0.017^{+0.028}_{-0.011}$ & &&
$0.5^{+0.9}_{-0.4}\times 10^{-6}$ &
\\ \hline\hline
& $\Upsilon(3S)\to J/\psi+\eta_c$    &&  &
$0.054^{+0.079}_{-0.033}$& &&
 $2.7^{+3.9}_{-1.6}\times 10^{-6}$ & &&$\Upsilon(4S)\to J/\psi+\eta_c$  & &&
 $0.031^{+0.046}_{-0.019}$ & &&
$1.5^{+2.2}_{-0.9}\times 10^{-9}$ & \\ \hline & $\Upsilon(3S)\to
J/\psi+\eta_c^\prime$ && & $0.027^{+0.043}_{-0.018}$ & &&
$1.3^{+2.1}_{-0.9}\times 10^{-6}$ & && $\Upsilon(4S)\to
J/\psi+\eta_c^\prime$& && $0.015^{+0.025}_{-0.010}$ & &&
$0.7^{+1.2}_{-0.5}\times 10^{-9}$ & \\
\hline & $\Upsilon(3S)\to \psi^\prime+\eta_c$    &&  &
$0.024^{+0.034}_{-0.014}$ & && $1.2^{+1.7}_{-0.7}\times 10^{-6}$ &
&&$\Upsilon(4S)\to \psi^\prime+\eta_c$ & &&
$0.014^{+0.019}_{-0.008}$ & && $0.7^{+1.0}_{-0.4}\times 10^{-9}$ &
\\ \hline & $\Upsilon(3S)\to \psi^\prime+\eta_c^\prime$    &&  & $0.011^{+0.018}_{-0.007}$ & &&
$0.6^{+0.9}_{-0.4}\times 10^{-6}$ & &&$\Upsilon(4S)\to
\psi^\prime+\eta_c^\prime$ & && $0.007^{+0.010}_{-0.004}$ & &&
$0.3^{+0.5}_{-0.2}\times 10^{-9}$ &
\\ \hline
\end{tabular}
\end{center} }
\end{table}

Our predictions to the partial widths and branching ratios for all
decay channels are listed in
Table~\ref{width:B:Upsilon:double:charmonium}.  One clearly sees
that the branching fractions for all decay processes (except for
$\Upsilon(4S)$) are about $10^{-6}$,   which are perfectly
compatible with the measured inclusive $J/\psi$ production rates
from $\Upsilon(nS)$ decay,
Eq.~(\ref{inclusive:Psi:production:ups}).  It is interesting to
note that our hadronic decay processes  have even smaller
branching ratios than the radiative decay $\Upsilon\to
\eta_c\gamma$ (${\cal B}\approx 3\times
10^{-5}$)~\cite{Guberina:1980xb}. This may be partly understood by
that $\Gamma[\Upsilon\to \eta_c\gamma]/\Gamma[\Upsilon\to
e^+e^-]\sim {\alpha_s^4\over\alpha}{m_c^2\over m_b^2} v_c^3$,
which has a milder $1/m_b^2$ scaling behavior compared to the
$1/m_b^6$ suppression in our processes, as manifested in
Eqs.~(\ref{scaling:bv:gamma}) and (\ref{scaling:bv:ggg}).

Between 2000 and 2003, CLEOIII has recorded about 20 million, 10
million and 5 million decays of $\Upsilon(1S)$, $\Upsilon(2S)$ and
$\Upsilon(3S)$, respectively~\cite{Briere:2004ug}.  So there
should be  a few to tens of produced events for each double
charmonium mode. Unfortunately, because the cleanest way of
tagging $J/\psi$ is through the dimuon mode, only 6\% fraction of
the produced events can be reconstructed.  Further taking into
account the  acceptance and efficiency to detect $\mu$,  it seems
rather difficult to observe these double charmonium production
events based on the existing CLEOIII data sample.  By contrast,
the high luminosity $e^+e^-$ colliders such as Belle and $BABAR$
have already collected a enormous amount of data at $\Upsilon(4S)$
peak. If they could dedicate some significant period of run at the
lower $\Upsilon$ resonances, it is feasible for them to discover
these decay channels unambiguously. Needless to say, the discovery
potential is very promising  for the planned super-high-luminosity
$e^+e^-$ facility like Super $B$ factory.

It is important to understand the interference pattern among three
different amplitudes. In our case, the phase in each amplitude
manifests itself as short-distance effect arising from loop,  and
is perturbatively calculable.  Let us take $\Upsilon(1S)\to
J/\psi+\eta_c$ as an example.  Taking $\xi= 4m_c^2/M^2_\Upsilon
\approx 0.10$ and $\alpha_s=0.22$, we find from
(\ref{a3g:analytic}) and (\ref{agggamma:analytic})
\bqa
a_{3g}&=&  3.89\, e^{-i\,105^\circ}\,,\hspace{2.2 cm}
a_{gg\gamma}= 0.44\, e^{\,i\,24^\circ}\,.
\label{Upsilon1S:phase:struct}
\eqa
Curiously, the strong decay amplitude is almost orthogonal to the
electromagnetic amplitude, while the radiative decay amplitude is
almost in phase with the electromagnetic one.  It is also obvious
to see that the strong decay amplitude has the most prominent
strength, the electromagnetic one the next, and the radiative
decay amplitude the least.

In digression, it may be instructive to know the relative
strengths of three different channels in inclusive $\Upsilon$
decay. From the following experimental inputs:
\bqa R &=& {\Gamma[\Upsilon\to \gamma^* \to X]\over
\Gamma[\Upsilon
\to \mu^+\mu^-] } = 3.56 \pm 0.07,\hspace{2.5 cm}\textrm{\cite{Ammar:1997sk}},\nn \\
R_\mu &=& {\Gamma[\Upsilon\to ggg]\over \Gamma[\Upsilon \to
\mu^+\mu^-]
} = 39.11\pm 0.4, \hspace{2.9 cm}\textrm{\cite{Eidelman:2004wy}}, \nn \\
R_\gamma &=& {\Gamma[\Upsilon\to gg\gamma]\over \Gamma[\Upsilon
\to ggg] } = 0.027\pm 0.003,\hspace{2.87 cm}\textrm{\cite{Besson:2005jv}},
\eqa
we can infer
\bqa {\cal B}[\Upsilon\to ggg]: {\cal B}[\Upsilon\to \gamma^*\to
X]:{\cal B}[\Upsilon\to gg\gamma] &= & 82.7\%:7.5\%:2.2\%\,,
\label{Br:inclusive:ratio} \eqa
where these three branching ratios sum up to $1-\sum{\cal
B}[\Upsilon\to l^+l^-]=92.5\% $, as they should~\footnote{We have
not included the contribution from the radiative transition
$\Upsilon\to \eta_b\gamma$, which has a completely negligible
branching ratio.}.  A very simple expectation is that each
amplitude in an exclusive process scales with the corresponding
$\sqrt{{\cal B}_{\rm incl}}$. The relative strengths of three
amplitudes in (\ref{Upsilon1S:phase:struct}) roughly respect this
scaling rule if one assumes  $K=1$. Nevertheless, the truly
important point is that, the orders of strengths of three
amplitudes are same for both inclusive and exclusive decays.

We can gain more intuition about the interference pattern by
examining the individual contribution to the partial width. Had we
retained only $a_\gamma$ in (\ref{Partial:Width}), the partial
width for $\Upsilon(1S)\to J/\psi+\eta_c$  would be only $0.065$
eV. If we kept $a_{3g}$ only,  the width would instead be $0.204$
eV. If we include both $a_\gamma$ and $a_{3g}$ but discard
$a_{gg\gamma}$, the width would become $0.210$ eV,  which is
rather close to the full answer listed in
Table~\ref{width:B:Upsilon:double:charmonium}, $0.208$ eV.  This
numerical exercise clearly corroborates our expectation about the
relative importance of these three different channels.

The phase structures in (\ref{Upsilon1S:phase:struct}) also hold
for other decay channels of $\Upsilon(nS)$ to double charmonium.
We take $\Upsilon(4S)\to J/\psi+\eta_c$ as second example to
verify this point.  Taking $\xi=4m_c^2/M^2_{\Upsilon(4S)} \approx
0.08$, we obtain
\bqa
a_{3g}&=& 4.20\, e^{-i\,102^\circ}\,,\hspace{2.2 cm}
a_{gg\gamma}= 0.52\, e^{\,i\,20^\circ}\,.
\label{Upsilon4S:phase:struct}
\eqa

It has been of great interest to decipher the interference pattern
between the strong and electromagnetic amplitude in $J/\psi$
decays.  The relative phase between $3g$ and $\gamma$ amplitude in
$J/\psi\to PV$ has  been determined to be around $-(106\pm
10)^\circ$~\cite{Baltrusaitis:1984rz,Coffman:1988ve,Jousset:1988ni,
LopezCastro:1994xw,Suzuki:1998ea,Achasov:2001wy}. This is
surprisingly close to our finding in $\Upsilon$ decay.   Suzuki
has argued that the large relative phase in $J/\psi$ decay must
arise from long-distance rescattering effect, and emphasized that
it is impossible for the perturbative quark-gluon process to
generate it~\cite{Suzuki:1998ea}.  However, our calculation
provides an explicit counterexample against his claim, showing
that the short-distance contribution alone suffices to generating
such a large relative phase.

It is worth mentioning that some years ago, Gerard and Weyers
argued there should be universal orthogonality between strong and
electromagnetic amplitude for each $J/\psi$ exclusive decay
mode~\cite{Gerard:1999uf}.  This assertion may seem to be backed
by numerous phenomenological evidences~\footnote{Besides the
$1^-0^-$ mode, other two-body decays of $J/\psi$ seem to also have
a nearly orthogonal relative phase between $a_\gamma$ and
$a_{3g}$, such as
$0^-0^-$~\cite{LopezCastro:1994xw,Suzuki:1999nb},
$1^-1^-$~\cite{Kopke:1988cs,LopezCastro:1994xw,Suzuki:1999nb},
$1^+0^-$~\cite{Suzuki:2001fs} and
$N\overline{N}$~\cite{LopezCastro:1994xw,Baldini:1998en}. Moreover
in $\psi^\prime$ decays, the $1^-0^-$~\cite{Wang:2003hy} and
$0^-0^-$ mode~\cite{Yuan:2003hj,Dobbs:2006fj} seem also compatible
with a large relative phase.}.  They have attributed this
orthogonality simply to the orthogonality of gluonic and one
photon states. Inspecting their arguments carefully, one finds
that they only prove the incoherence between three-gluon and
single-photon decays at {\it inclusive} level, whose validity
crucially relies on summing over all possible decay channels.
Since there is no room for such a summation for exclusive $J/\psi$
decay,  there is no any simple reason to believe why strong decay
amplitude should be orthogonal to the electromagnetic amplitude
channel by channel.

Because their reasoning is based on rather general ground, one may
test it in $\Upsilon$ exclusive decay.  As a matter of fact,  we
can directly present a counterexample.  Imagine a fictitious world
with an extremely heavy $b$ quark, say $m_b \sim M_{\rm Plank}$,
but with an ordinary charm quark.  For the would-be $\Upsilon$
decay to $J/\psi+\eta_c$, we then find from
(\ref{Re:f:asymptotic}) and (\ref{Im:f:asymptotic}) that the phase
of $f$ is very close to zero, so is the relative phase between
$a_{3g}$ and $a_\gamma$.

One may wonder why Gerard and Weyers's assertion seems to enjoy
considerable success when applied to $J/\psi$ decays, even though
it looks theoretically ungrounded. One possible explanation is
that, due to some specific dynamics,  the relative strength and
phase between electromagnetic and strong amplitudes are roughly
identical for each $J/\psi$ exclusive decay mode, preserving the
same pattern as in the inclusive decay. This approximate scaling
between exclusive and inclusive channels is exemplified in the
discussion following (\ref{Br:inclusive:ratio}).  This pattern
does not necessarily hold for other vector quarkonium decays.

It is straightforward to see that,  the approximate $-90^\circ$
phase between strong and electromagnetic amplitude in our process
is simply a consequence of the not-too-tiny mass ratio
$m_c^2/m_b^2\approx 0.1$ and the opposite electric charges of $c$
and $b$ (see left panel of Fig.~\ref{plot:f:g} and
(\ref{a3g:analytic})). It may seem to be a marvellous coincidence
that the relative phase determined in our case is very close to
that in $J/\psi\to PV$,  especially regarding that the latter
process should be largely dictated by nonperturbative
long-distance dynamics.  We don't  know exactly  which
nonperturbative mechanism should be responsible for the universal
orthogonal phase in various $J/\psi$ decay modes.  It is fun to
notice that, however, in the constituent quark model, the masses
of $u$, $d$ and $s$ quarks are several hundreds of MeV,
consequently $m^2_{u,d,s}/m_c^2\approx m_c^2/m_b^2$, so our
formalism seems to be able to explain the nearly orthogonal phase
in $J/\psi\to PV$ entirely within the short-distance quark-gluon
picture.

Lastly we stress that the phases determined in
(\ref{Upsilon1S:phase:struct}) and (\ref{Upsilon4S:phase:struct})
are subject to large uncertainties.  Since they are determined
only at the lowest-order accuracy,  it is conceivable that they
may receive large modifications by including radiative and
relativistic corrections.  Moreover, for simplicity we have
assumed the radiative correction to electromagnetic amplitude does
not introduce  an imaginary part.  One should realize this is just
an (decent though) approximation.  Despite this alertness, we
still expect the qualitative feature, {\it i.e.}, the large
relative phase can withstand all these uncertainties.

\subsection{Continuum-resonance interference for double
charmonium production}

For a given final state in $e^+e^-$ annihilation experiment near a
vector meson resonance, it is always produced via two inseparable
mechanisms-- resonant decay and continuum production. A rough
indicator about the relative strength of resonant electromagnetic
amplitude to the continuum amplitude is characterized by $3\,{\cal
B}_{e^+e^-}/\alpha$. For the first four $\Upsilon$ resonances,
this factor is 10.2, 7.9, 8.9 and 0.0055 respectively. Therefore,
for the three lower $\Upsilon$ resonances, the $J/\psi+\eta_c$
production are dominated by the resonant decay, whereas for the
$\Upsilon(4S)$, which has a width about three orders of magnitude
broader,  one expects that the continuum contribution plays an
overwhelmingly important role.

\begin{table}[tb]
\caption{The Breit-Wigner, continuum and full cross sections (in
units of fb) for $e^+e^-\to J/\psi+\eta_c$ at various
$\Upsilon(nS)$ resonances. All the input parameters are the same
as in Section~\ref{excl:dec:ups:2swave:charmonium} except
$\alpha_s$ is fixed to be 0.22.} \label{peak:xsection:cont:BW}
\begin{center}
\begin{tabular}{ccc ccc ccc ccc ccc}
\hline  & $\sqrt{s}$  (GeV) & && $\sigma_{\rm BW}$  & &&
$\sigma_{\rm cont}$  & && $\sigma_{\rm full}$  &
 \\ \hline
 & $9.460$ & &&
 $15678$ & && 47.1  & && 14158 &
 \\ \hline & 10.023& && 7165 & &&
32.7& &&  6317 &  \\
\hline  & 10.355 & && 7948 & && 26.4 & && 7141 &
\\ \hline & $10.579$& && 0.0026 & && 22.9
& && 22.5 &
\\ \hline
\end{tabular}
\end{center}
\end{table}

We are interested to know the impact of the resonance-continuum
interference on the observed cross sections.  Assuming $a_\gamma$
and $a_c$ differ by a Breit-Wigner propagator,  one can express
the full cross section near $\Upsilon$ peak as
\bqa \sigma_{\rm full}[e^+e^- \to  J/\psi+\eta_c] &=&
\sigma_{\mu^+\mu^-} \,{2^{20}\,\pi^2\,e_c^2\alpha_s^2 \over 9 }
{|{\bf P}|^3 \over s^{9/2}} \, \psi^2_{J/\psi}(0)\,
\psi^2_{\eta_c}(0)
\\
&\times & \left|K+ {3 \alpha^{-1}\,\sqrt{s}\:\Gamma_{e^+e^-}\over
s-M_\Upsilon^2+iM_\Upsilon\Gamma_{\rm tot}} \,
(K+a_{3g}+a_{gg\gamma})\right|^2, \nn \eqa
where $\Gamma_{e^+e^-}$ and $\Gamma_{\rm tot}$ are the electric
and total width of $\Upsilon$. If the continuum term is dropped,
this formula then reduces to the standard Breit-Wigner form:
\bqa \sigma_{\rm BW}[e^+e^-\to \Upsilon \to J/\psi+\eta_c] &=& {12
\pi \,\Gamma_{e^+e^-}\,\Gamma[\Upsilon \to J/\psi+\eta_c] \over
\left(s-M_\Upsilon^2\right)^2+ M_\Upsilon^2 \,\Gamma_{\rm
tot}^2}\,. \eqa

In Table~\ref{peak:xsection:cont:BW} we have enumerated  various
contributions to the $J/\psi+\eta_c$ cross sections at
$\Upsilon(nS)$ peaks. One can clearly see the inclusion of the
continuum contribution will reduce the peak cross sections by
about 10\% for the first three $\Upsilon$ states, whereas
including the resonant contribution will reduce the continuum
cross section by about 2\% for $\Upsilon(4S)$.  This destructive
interference can be attributed to the approximate $180^\circ$
relative phase between $a_{3g}$ and $a_c$.

The interference with continuum contribution also slightly
distorts the Breit-Wigner shape of the production cross sections
for the first three $\Upsilon$ resonances.  However, one has to
bear in mind that, for a thorough analysis,  one has to carefully
take the beam spread and radiative corrections into account, which
requires lots of extra work and we leave them to the
experimentalists.

\begin{figure}[tb]
\begin{center}
\includegraphics[scale=0.5]{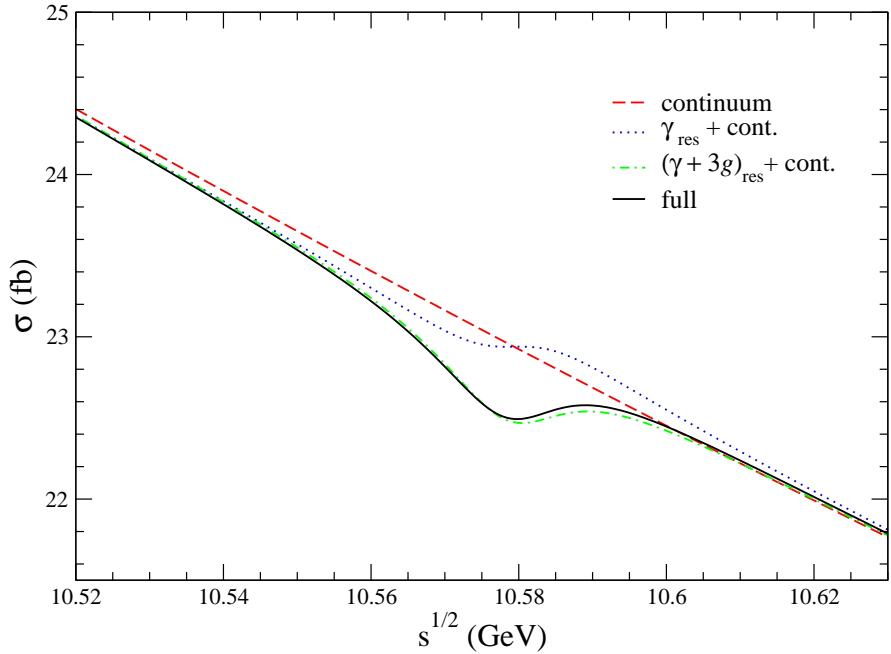}
\caption{The line shape of $e^+e^-\to J/\psi+\eta_c$ near
$\sqrt{s}=M_{\Upsilon(4S)}$. \label{Upsilon4S:line:shape}}
\end{center}
\end{figure}

Thus far, the measured double charmonium production in $B$
factories has been assumed to be entirely initiated by the
continuum process, as represented in
(\ref{double:charmonim:conti:tree}). Experimentally, the resonant
decay, despite its small magnitude, is  encapsulated in the
observed cross sections.  It is interesting to know how the line
shape of $J/\psi+\eta_c$ near $\Upsilon(4S)$ peak would be
affected by including this contribution.  In
Fig.~\ref{Upsilon4S:line:shape}, we have shown the various line
shapes,  with the contributions from several different sources
juxtaposed.   An interesting feature is that a dip is developed
right on the $\Upsilon(4S)$ peak, which is again due to the
destructive interference between the resonant strong decay and
continuum amplitudes.  Furthermore, we are reassured again that
the radiative decay amplitude is unimportant. It will be great if
someday experimentalists can do an energy scan and pin down this
dip structure. To achieve this goal, the cross section must be
measured very precisely, of course a very challenging task.  We
finally remark that, due to the aforementioned destructive
interference, the true continuum cross sections should be slightly
larger than the values quoted in (\ref{Babar:doub:charmon:meas}),
which are in fact the full cross sections measured experimentally.

\section{Summary and Outlook}
\label{summary}

In this work, we have performed a systematic study on $\Upsilon$
exclusive decays to vector plus pseudoscalar charmonium in NRQCD
factorization framework.  These exclusive decay modes can proceed
via three-gluon, one-photon and two-gluon-one-photon, each of
which has been thoroughly analyzed.  The relative phases among
these amplitudes naturally arise as a consequence of the
short-distance loop contribution.   A particularly interesting
finding is that the relative phase between strong and
electromagnetic amplitude is nearly orthogonal,  which is the same
as that in various $J/\psi$ decay modes.

The typical branching fractions of these decays  are predicted to
be of order $10^{-6}$ for the low-lying $\Upsilon(nS)$ states
($n=1,2,3$).  Future dedicated high-luminosity $e^+e^-$
facilities, {\it e.g.} Super $B$ experiment,  should be able to
discover these decay channels readily.

We have also investigated the impact of the continuum-resonance
interference on the $J/\psi+\eta_c$ production cross sections at
different $\Upsilon$ peaks.  We find this interference will reduce
the peak cross sections for the first three $\Upsilon$ states by
about 10\%.   We predict there is a small dip in the line shape on
the $\Upsilon(4S)$ peak.  The current experiments are too rough to
discern this delicate structure,  perhaps the future Super $B$
experiment can verify this prediction.

A natural extension of this work is to investigate other exclusive
double charmonium production processes from $\Upsilon$ decay. For
example, $\Upsilon\to \chi_{cJ}\,J/\psi$ are particularly
interesting channels to study,  since the inclusive bounds for
$\Upsilon\to \chi_{cJ}+X$ have already been experimentally
available~\cite{Briere:2004ug}.  Besides these double charmonium
decay modes, one may also be tempted to apply the same formalism
developed in this work to the processes $\Upsilon(J/\psi)\to
PV$~\cite{Jia:prepare}.  For the scarcity of theoretical
investigations to these decay modes from the angle of pQCD, this
study will offer us something worthwhile learning.  Although it
will no longer be as theoretically well grounded as the processes
considered in this work,   it should be viewed as an approach
rooted in the time-tested constituent quark model, which has
witnessed many phenomenological successes over years.

\acknowledgments

I am indebted to Chang-Zheng Yuan for comments on the manuscript.
This work is supported in part by National Natural Science
Foundation of China under Grant No.~10605031.

\appendix
\section{Deriving analytical expression for $f$}\label{f:analytic}

In this Appendix we illustrate how to simplify $f$ effectively, so
that we can obtain their analytical expressions.  Repeatedly using
kinematical relations stemming from the constraint
$k_1+k_2={Q\over 2}$, plus fractional decomposition,  we can
reduce each $f_i$ in (\ref{3g:f1}), (\ref{3g:f2}) and
(\ref{3g:f3}) into the sum of two-point tensor and three-point
scalar integrals:
\bqa f_1(\xi)&=& \int\! {d^4 k_1\over i\pi^2} \left\{ {2m_b^2-3m^2_c
\over k_1^2 k_2^2 (k_1^2-k_1\cdot P)} + {k_1\cdot(3 Q-P) \over
m_b^2 }\left[ {1\over k_1^2  k_2^2}- {1\over k_1^2 (k_2^2-m_b^2)}
\right] \right.
\\
& + &  \left.  {k_1\cdot [(2-\xi)Q-P] \over m_b^2}
 \left[
{1\over (k_2^2-m_b^2)(k_1^2-k_1\cdot P)} - {1\over k_2^2
(k_1^2-k_1\cdot P)} \right] \right\} \,,
\nn \\
f_2(\xi) &=& \int\! {d^4 k_1\over i\pi^2} \left\{ {m_b^2-3m_c^2
\over k_1^2 k_2^2 (k_1^2-k_1\cdot P)} + {k_2\cdot P \over
m_b^2}\left[ {1\over k_1^2 k_2^2} -{1\over (k_1^2-m_b^2) k_2^2}
\right]  \right.
\\
& + &  \left.  {k_2\cdot (P-\xi Q) \over m_b^2}
 \left[ {1\over (k_1^2-m_b^2) (k_1^2-k_1\cdot P)}-{1\over k_1^2
(k_1^2-k_1\cdot P)} \right] \right\}\,,
\nn \\
 f_{3}(\xi) &=& -2\int\!
{d^4 k_1\over i\pi^2} \left\{ {m_b^2 -2 m_c^2 \over k_1^2 k_2^2
(k_1^2-k_1\cdot P)}+ {2 m_c^2 \over (k_1^2-m_b^2)(k_2^2-m_b^2)
(k_1^2-k_1\cdot P)} \right.
\\
&+& \left. {1\over k_1^2 k_2^2} -
 {1\over (k_1^2-m_b^2) (k_2^2-m_b^2)}+{1\over (k_1^2-m_b^2)(k_1^2-k_1\cdot P)}-
 {1\over k_1^2 (k_1^2-k_1\cdot P)}
\right\}, \nn
 \eqa
where $\xi= m_c^2/m_b^2$. While the two-point functions can be
trivially handled, working out the three-point scalar integrals is
more laborious but still straightforward.  Here we just give their
analytic forms:
\bqa C_1(\xi) &=& \int\! {d^4 k_1 \over i\pi^2}\, {m_b^2 \over k_1^2
\,k_2^2\,(k_1^2-k_1\cdot P)}
\nn \\
&=& -{1\over \beta} \left\{2\,{\rm tanh}^{-1}\beta\,\ln \xi +2\,{\rm
Li}_2\left[{1-\beta\over 2}\right] -2\,{\rm
Li}_2\left[{1+\beta\over 2}\right] \right.
\nn \\
&+& \left. {\rm Li}_2\left[{\beta-1\over \beta+1}\right] - {\rm
Li}_2\left[{\beta+1\over \beta-1}\right]+ 2\pi i\,{\rm
tanh}^{-1}\beta \right\}\,. \label{3point:1st:kind}
\\
 C_2(\xi) &=& \int\! {d^4 k_1\over i\pi^2 }\,
{m_b^2 \over (k_1^2-m_b^2) \,(k_2^2- m_b^2)\,(k_1^2-k_1\cdot P)}
\nn \\
&=& -{1\over \beta} \left\{ {2 \pi \over 3}{\rm
tan}^{-1}[\sqrt{3}\beta] + 2\,{\rm tanh}^{-1}\beta\, \ln [1-3 \xi]
+{\rm Li}_2\left[{2\beta\over 1+\beta}\right] -{\rm
Li}_2\left[{2\beta\over \beta-1}\right] \right.
\nn \\
&+& {\rm Li}_2\left[{\beta(\beta+1)\over \beta-1}\right]  -{\rm
Li}_2\left[{\beta(1-\beta)\over 1+\beta}\right] +{\rm
Li}_2\left[{\beta(1+\beta)\over 2 (1-3 \xi)}\right] -{\rm
Li}_2\left[{\beta(\beta-1)\over 2 (1-3 \xi)}\right]
\nn \\
&+& {\rm Li}_2\left[-{\beta(1-\beta)^2\over 4 (1-3 \xi)}\right]
-{\rm Li}_2\left[{\beta(1+\beta)^2\over 4 (1-3 \xi)}\right]
\nn \\
&+& \left. 2\,{\rm Re}\left\{ {\rm Li}_2\left[-{(1+
i\,\sqrt{3})\beta \over 1-i\,\sqrt{3}\,\beta } \right]-{\rm
Li}_2\left[{(1+ i\,\sqrt{3})\beta \over 1+i\,\sqrt{3}\,\beta }
\right]\right\} \right\}\,. \label{3point:2nd:kind} \eqa
It may be worth mentioning that if the well-known master formula
for massive three-point scalar integral ({\it i.e.}, equation
(5.6) in \cite{'tHooft:1978xw}) is employed, one seems unable to
obtain the correct expression for $C_2$.  To be specific, using
that formula would render $C_2({1\over 4})=0$, which diametrically
conflicts with the true value $4 \ln2- 2 \pi/\sqrt{3}$. One can
check our result is correct.

We now can express $f_i$ as follows:
\bqa f_1(\xi)&=& (2-3\xi)\,C_1(\xi)+{5\over
2}+2(3-2\xi)\left\{{1\over 1-\beta}\ln\left[{1+\beta\over
2}\right]+ {1\over
1+\beta}\ln\left[{1-\beta\over 2}\right]\right\} \nn \\
&-&   2 (1+\xi) \left\{ {1\over (1-\beta)^2}\ln\left[{1+\beta\over
2}\right]+ {1\over (1+\beta)^2}\ln\left[{1-\beta\over 2}\right]
+{1\over 4 \xi}\right\}+{5\,i\,\pi\over 2}\,,
\\
f_2(\xi)&=& (1-3\xi)\,C_1(\xi)+{1\over 2}+2(1-2\xi)\left\{{1\over
1-\beta}\ln\left[{1+\beta\over 2}\right]+ {1\over
1+\beta}\ln\left[{1-\beta\over 2}\right]\right\} \nn \\
&-&  2 \xi \left\{ {1\over (1-\beta)^2}\ln\left[{1+\beta\over
2}\right]+ {1\over (1+\beta)^2}\ln\left[{1-\beta\over 2}\right]
+{1\over 4 \xi}\right\}+{i\,\pi\over 2}\,,
\\
f_3(\xi)&=& -2(1-2\xi)\,C_1(\xi)- 4\xi\,C_2(\xi)- 4\left\{{1\over
1-\beta}\ln\left[{1+\beta\over 2}\right]+ {1\over
1+\beta}\ln\left[{1-\beta\over 2}\right] \right\} \nn \\
&-&  {2\,\pi\over \sqrt{3}} -2\,i\,\pi\,.
\eqa
Adding these three functions together then reproduces
(\ref{re:f:tot}) and (\ref{im:f:tot}).

\section{Deriving analytical expression for $g$}\label{g:analytic}

In this Appendix we illustrate how to reduce the one-loop
four-point function in (\ref{gggamma:def:g}) to the sum of simpler
two- and three-point scalar integrals.  With the aid of the
kinematical identities arising from the constraint
$k_1+k_2=\tilde{P}$, we can disentangle this integral into three
pieces:
\bqa g(\xi)&=& g_1(\xi)+g_2(\xi)+g_3(\xi)\,, \eqa
where $\xi= m_c^2/m_b^2$, and
\bqa g_1(\xi) &=& {1 \over 2}\int\! {d^4 k_1 \over i\pi^2} {1\over
k_1^2}\left[{1\over k_1^2-k_1\cdot Q}-{1\over k_2^2-k_2\cdot Q}
\right]\,,
\\
g_2(\xi) &=& {2m_c^2\over m_b^2} \int\! {d^4 k_1 \over i\pi^2}
\,\left[{1\over k_1^2}-{1\over k_1^2-k_1\cdot Q} \right]\,{1\over
k_2^2-k_2\cdot Q}
\nn \\
&+& 2 \int\! {d^4 k_1 \over i\pi^2 } \, {m_c^2 \over k_1^2
(k_1^2-k_1\cdot Q) (k_2^2-k_2\cdot Q)}\,,
\\
g_3(\xi) &=& {2(m_b^2-4 m_c^2)\over m_b^2}\int\! {d^4 k_1 \over
i\pi^2 } \left[ {m_c^2 \over k_1^2 k_2^2 (k_1^2-k_1\cdot Q)}+
{m_b^2-m_c^2 \over k_1^2 (k_1^2-k_1\cdot Q) (k_2^2-k_2\cdot
Q)}\right]. \eqa

Here we give the analytical expressions of two needed scalar
3-point integrals:
\bqa \tilde{C}_1(\xi) &=& \int\! {d^4 k_1 \over i\pi^2}\, {m_b^2
\over k_1^2 \,k_2^2\,(k_1^2-k_1\cdot Q)}
\nn \\
&=& -{1\over 2\beta} \left\{4\,{\rm tanh}^{-1}\beta\,\ln [2\xi]
+2\,{\rm Li}_2[-\beta] -2\,{\rm Li}_2[\beta]+  {\rm
Li}_2\left[{\beta-1\over \beta+1}\right] \right.
\nn \\
&-& \left. {\rm Li}_2\left[{1+\beta\over \beta-1}\right]+ {\rm
Li}_2\left[{2\,\beta\over (1+\beta)^2}\right] - {\rm
Li}_2\left[-{2\,\beta\over (1-\beta)^2}\right] +2\pi i\,{\rm
tanh}^{-1}\beta \right\}\,,
\\
\tilde{C}_2(\xi) &=& \int\! {d^4 k_1 \over i\pi^2}\, {m_b^2 \over
k_1^2 \,(k_1^2-k_1\cdot Q)\,(k_2^2-k_2\cdot Q)}
\nn \\
&=& -{1\over 2\beta} \left\{ {\rm Li}_2[\beta]- {\rm
Li}_2[-\beta]+2\,{\rm Re} \left\{ {\rm
Li}_2\left[{(1+\beta)^2+4i\sqrt{\xi(1-\xi)}\over 4(1-2\xi)}\right]
\right.\right. \nn \\
&-& {\rm Li}_2\left[{(1-\beta)^2+4i\sqrt{\xi(1-\xi)}\over
4(1-2\xi)}\right]+ {\rm
Li}_2\left[-{\beta(1-\beta)^2+4i\beta\sqrt{\xi(1-\xi)}\over
4(1-2\xi)}\right]
 \nn \\
&-& \left.\left.
 {\rm
Li}_2\left[{\beta(1+\beta)^2+4i\beta\sqrt{\xi(1-\xi)}\over
4(1-2\xi)}\right] \right\}\right\}\,.
 \label{3point:1st:kind} \eqa

Therefore we have
\bqa g_1(\xi) &=& {1-2\xi \over 1-4\xi}\, \ln[2-4\xi]\,,
\\
g_2(u) &=& 4\xi\left( \sqrt{1-\xi \over \xi}\, \tan^{-1}\sqrt{\xi
\over 1-\xi}-
 {1-2 \xi \over 1-4 \xi}\, \ln[2- 4 \xi] \right)+ 2 \xi\,\tilde{C}_2(u)\,,
\\
g_3(\xi) &=& 2(1-4 \xi)[ \xi\,
\tilde{C}_1(\xi)+(1-\xi)\,\tilde{C}_2(\xi)]\,.\eqa
One then readily reproduces the analytic results shown in
(\ref{re:g:tot}) and (\ref{im:g:tot}).



\begin{thebibliography}{99}


\bibitem{Kopke:1988cs}
  L.~Kopke and N.~Wermes,
  Phys.\ Rept.\  {\bf 174}, 67 (1989).

\bibitem{Brambilla:2004wf}
N.~Brambilla {\it et al}, CERN-2005-005 [arXiv:hep-ph/0412158].


\bibitem{Bodwin:1994jh}
  G.~T.~Bodwin, E.~Braaten and G.~P.~Lepage,
  Phys.\ Rev.\  D {\bf 51}, 1125 (1995)
  [Erratum-ibid.\  D {\bf 55}, 5853 (1997)]
  [arXiv:hep-ph/9407339].


\bibitem{Braguta:2005gw}
  V.~V.~Braguta, A.~K.~Likhoded and A.~V.~Luchinsky,
  Phys.\ Rev.\ D {\bf 72}, 094018 (2005)
  [arXiv:hep-ph/0506009].

\bibitem{Jia:2006rx}
  Y.~Jia,
  arXiv:hep-ph/0611130.


\bibitem{Maschmann:1989ai}
  W.~S.~Maschmann {\it et al.}  [Crystal Ball Collaboration],
  Z.\ Phys.\  C {\bf 46}, 555 (1990).


\bibitem{Abe:2001za}
  K.~Abe {\it et al.}  [BELLE Collaboration],
  Phys.\ Rev.\ Lett.\  {\bf 88}, 052001 (2002)
  [arXiv:hep-ex/0110012].


\bibitem{Briere:2004ug}
  R.~A.~Briere {\it et al.}  [CLEO Collaboration],
  Phys.\ Rev.\  D {\bf 70}, 072001 (2004)
  [arXiv:hep-ex/0407030].


\bibitem{Brodsky:1981kj}
  S.~J.~Brodsky and G.~P.~Lepage,
  Phys.\ Rev.\  D {\bf 24}, 2848 (1981).

\bibitem{Chernyak:1983ej}
  V.~L.~Chernyak and A.~R.~Zhitnitsky,
  Phys.\ Rept.\  {\bf 112}, 173 (1984);


\bibitem{Abe:2002rb}
  K.~Abe {\it et al.}  [Belle Collaboration],
  Phys.\ Rev.\ Lett.\  {\bf 89}, 142001 (2002)
  [arXiv:hep-ex/0205104].


\bibitem{Baltrusaitis:1984rz}
  R.~M.~Baltrusaitis {\it et al.}  [MARK-III Collaboration],
  Phys.\ Rev.\  D {\bf 32}, 2883 (1985).


\bibitem{Coffman:1988ve}
  D.~Coffman {\it et al.}  [MARK-III Collaboration],
  Phys.\ Rev.\  D {\bf 38}, 2695 (1988)
  [Erratum-ibid.\  D {\bf 40}, 3788 (1989)].


\bibitem{Jousset:1988ni}
  J.~Jousset {\it et al.}  [DM2 Collaboration],
  Phys.\ Rev.\  D {\bf 41}, 1389 (1990).


\bibitem{LopezCastro:1994xw}
  G.~Lopez Castro, J.~L.~Lucio M. and J.~Pestieau,
  AIP\ Conf.\ Proc. {\bf 342}, 441 (1995)
[arXiv:hep-ph/9902300].


\bibitem{Suzuki:1998ea}
  M.~Suzuki,
  Phys.\ Rev.\  D {\bf 57}, 5717 (1998)
  [arXiv:hep-ph/9801284].


\bibitem{Achasov:2001wy}
  N.~N.~Achasov,
  AIP Conf.\ Proc.\  {\bf 619}, 649 (2002)
  [arXiv:hep-ph/0110057].


\bibitem{Mo:2006cy}
For a recent review on the status of $\rho\pi$ puzzle, see
X.~H.~Mo, C.~Z.~Yuan and P.~Wang,
  arXiv:hep-ph/0611214.



\bibitem{Li:2007pb}
  T.~Li, S.~M.~Zhao and X.~Q.~Li,
  arXiv:0705.1195 [hep-ph].


\bibitem{Guberina:1980xb}
  B.~Guberina and J.~H.~Kuhn,
  Lett.\ Nuovo Cim.\  {\bf 32}, 295 (1981).




\bibitem{Braaten:2002fi}
  E.~Braaten and J.~Lee,
  Phys.\ Rev.\  D {\bf 67}, 054007 (2003)
  [Erratum-ibid.\  D {\bf 72}, 099901 (2005)]
  [arXiv:hep-ph/0211085].

\bibitem{Liu:2002wq}
  K.~Y.~Liu, Z.~G.~He and K.~T.~Chao,
  Phys.\ Lett.\  B {\bf 557}, 45 (2003)
  [arXiv:hep-ph/0211181].

\bibitem{Bodwin:2002fk}
  G.~T.~Bodwin, J.~Lee and E.~Braaten,
  Phys.\ Rev.\ Lett.\  {\bf 90}, 162001 (2003)
  [arXiv:hep-ph/0212181].

\bibitem{Bodwin:2002kk}
  G.~T.~Bodwin, J.~Lee and E.~Braaten,
  Phys.\ Rev.\  D {\bf 67}, 054023 (2003)
  [Erratum-ibid.\  D {\bf 72}, 099904 (2005)]
  [arXiv:hep-ph/0212352].



\bibitem{Hagiwara:2003cw}
  K.~Hagiwara, E.~Kou and C.~F.~Qiao,
  Phys.\ Lett.\  B {\bf 570}, 39 (2003)
  [arXiv:hep-ph/0305102].

\bibitem{Ma:2004qf}
  J.~P.~Ma and Z.~G.~Si,
  Phys.\ Rev.\  D {\bf 70}, 074007 (2004)
  [arXiv:hep-ph/0405111].


\bibitem{Bondar:2004sv}
  A.~E.~Bondar and V.~L.~Chernyak,
  Phys.\ Lett.\  B {\bf 612}, 215 (2005)
  [arXiv:hep-ph/0412335].


\bibitem{Zhang:2005ch}
  Y.~J.~Zhang, Y.~j.~Gao and K.~T.~Chao,
  Phys.\ Rev.\ Lett.\  {\bf 96}, 092001 (2006)
  [arXiv:hep-ph/0506076].

\bibitem{Bodwin:2006dm}
  G.~T.~Bodwin, D.~Kang and J.~Lee,
  Phys.\ Rev.\  D {\bf 74}, 114028 (2006)
  [arXiv:hep-ph/0603185].

\bibitem{He:2007te}
  Z.~G.~He, Y.~Fan and K.~T.~Chao,
  Phys.\ Rev.\  D {\bf 75}, 074011 (2007)
  [arXiv:hep-ph/0702239].


\bibitem{Yao:2006px}
  W.~M.~Yao {\it et al.}  [Particle Data Group],
  J.\ Phys.\ G {\bf 33}, 1 (2006).


\bibitem{Abe:2004ww}
  K.~Abe {\it et al.}  [Belle Collaboration],
  Phys.\ Rev.\  D {\bf 70}, 071102 (2004)
  [arXiv:hep-ex/0407009].

\bibitem{Aubert:2005tj}
  B.~Aubert {\it et al.}  [BABAR Collaboration],
  Phys.\ Rev.\  D {\bf 72}, 031101 (2005)
  [arXiv:hep-ex/0506062].


\bibitem{Ammar:1997sk}
  R.~Ammar {\it et al.}  [CLEO Collaboration],
  Phys.\ Rev.\  D {\bf 57}, 1350 (1998)
  [arXiv:hep-ex/9707018].

\bibitem{Eidelman:2004wy}
  S.~Eidelman {\it et al.}  [Particle Data Group],
  Phys.\ Lett.\  B {\bf 592}, 1 (2004).



\bibitem{Besson:2005jv}
  D.~Besson {\it et al.}  [CLEO Collaboration],
  Phys.\ Rev.\  D {\bf 74}, 012003 (2006)
  [arXiv:hep-ex/0512061].



\bibitem{Gerard:1999uf}
  J.~M.~Gerard and J.~Weyers,
  Phys.\ Lett.\  B {\bf 462}, 324 (1999)
  [arXiv:hep-ph/9906357].


\bibitem{Suzuki:1999nb}
  M.~Suzuki,
  Phys.\ Rev.\  D {\bf 60}, 051501 (1999)
  [arXiv:hep-ph/9901327].


\bibitem{Suzuki:2001fs}
  M.~Suzuki,
  Phys.\ Rev.\  D {\bf 63}, 054021 (2001).


\bibitem{Baldini:1998en}
  R.~Baldini {\it et al.},
  Phys.\ Lett.\  B {\bf 444}, 111 (1998).

\bibitem{Wang:2003hy}
  P.~Wang, C.~Z.~Yuan and X.~H.~Mo,
  Phys.\ Rev.\  D {\bf 69}, 057502 (2004)
  [arXiv:hep-ph/0303144].


\bibitem{Yuan:2003hj}
  C.~Z.~Yuan, P.~Wang and X.~H.~Mo,
  Phys.\ Lett.\  B {\bf 567}, 73 (2003)
  [arXiv:hep-ph/0305259].


\bibitem{Dobbs:2006fj}
  S.~Dobbs {\it et al.}  [CLEO Collaboration],
  Phys.\ Rev.\  D {\bf 74}, 011105 (2006)
  [arXiv:hep-ex/0603020].


\bibitem{Jia:prepare}
  Y.~Jia, in preparation.


\bibitem{'tHooft:1978xw}
  G.~'t Hooft and M.~J.~G.~Veltman,
  Nucl.\ Phys.\ B {\bf 153}, 365 (1979).


\end{thebibliography}
\end{document}